Time-reversal symmetry breaking type II Weyl state in YbMnBi$_2$

Sergey Borisenko[1], Daniil Evtushinsky[1,2], Quinn Gibson[3], Alexander Yaresko[4], Timur Kim[5], M. N. Ali[3], Bernd Büchner[1,6], Moritz Hoesch[5], Robert J. Cava[3]

[1]*IFW-Dresden, Helmholtzstrasse 20, 01069 Dresden, Germany*

[2]*BESSY II, Helmholtz-Zentrum Berlin for Materials and Energy, Albert-Einstein-Strasse 15, 12489 Berlin, Germany*

[3] *Department of Chemistry, Princeton University, Princeton, New Jersey 08544, USA*

[4]*Max-Planck-Institute for Solid State Research, Heisenbergstrasse 1, 70569 Stuttgart, Germany*

[5] *Diamond Light Source, Harwell Campus, Didcot, OX11 0DE, United Kingdom*

[6]*Institute for Solid State Physics, TU Dresden, 01062 Dresden, Germany*

**Detection of Dirac, Majorana and Weyl fermions in real materials may significantly strengthen the bridge between high-energy and condensed-matter physics. While the presence of Dirac fermions is well established in graphene and topological insulators [1, 2], Majorana particles have been reported recently [3, 4] and evidence for Weyl fermions in non-centrosymmetric crystals has been found only a couple of months ago [5-11], the "magnetic" Weyl fermions are still elusive despite numerous theoretical predictions and intense experimental search [12-17]. In order to detect a time-reversal symmetry breaking Weyl state we designed two materials with Fermi velocities superior to that of graphene and present here the experimental evidence of the realization of such a state in one of them, YbMnBi$_2$. We model the time reversal symmetry breaking observed by magnetization measurements by a canted antiferromagnetic state and find a number of Weyl points both above and below the Fermi level. Using angle-resolved photoemission, we directly observe these latter Weyl points and a hallmark of the exotic state – the arc of the surface states which connects these points. Our results not only provide a fundamental link between the two areas of physics, but also demonstrate the practical way to design novel materials with exotic properties.**

Although famous as some of the most influential theories in particle physics, two nearly century old papers by Paul Dirac and Hermann Weyl were about an electron, as per both titles [18, 19]. To overcome the incompleteness of the previous theories, Dirac suggested a new description of the electron in terms of four-component wave functions which satisfied both quantum mechanics and relativity. Weyl has demonstrated that in the massless case the description can be more compact and needs only two-component wave function. Because all fermions of the Standard Model are Dirac fermions except possibly neutrinos (which are not a Weyl fermions because of finite mass), Weyl fermions escaped the detection until recently [5-11].

To observe Weyl fermions in three-dimensional (3D) condensed matter systems, the lifting of the degeneracy of the electronic bands is required, and singly degenerate bands should contain relativistic and massless states which touch in a single point. Only then will the Schrödinger equation be similar to a two-component Weyl equation and the electrons in this state will behave as Weyl fermions. These requirements can be realized in noncentrosymmetric and magnetic materials with strong spin-orbit interaction and many compounds have already been nominated to host this exotic state [5, 6, 12-17].

In order to satisfy all mentioned above criteria, we have selected a material that contains structural elements capable of generating the required elements of the electronic structure, focusing on magnetic systems with a center of inversion. The most simple and well-known "generator" of the Dirac-like massless dispersions is the 2D network of Bi atoms. This choice also guarantees a sizeable spin-orbit splitting. Another constraint was dictated by the necessity to stay close to the anti-ferromagnetic (AFM) order, as in Ref. 5, to avoid strongly magnetized domains which would complicate the detection of the state by ARPES. Finally, time-reversal symmetry (TRS) has still to be broken in order to ensure the lifting of the degeneracy. On the basis of these considerations we have grown single crystals of $EuMnBi_2$ (a previously reported compound [20]) and $YbMnBi_2$ (not a previously reported compound).

Our ARPES and magnetization measurements clearly demonstrate the time-reversal symmetry breaking in one of the materials, $YbMnBi_2$, and this is in excellent agreement with the calculations considering a canted antiferromagnetic order. The fully relativistic calculations reveal a number of Weyl points and predict $YbMnBi_2$ to be the first canted antiferromagnet hosting a magnetic Weyl state of the second type [21]. We detect the Weyl points as well as an extra state (Fermi arc) of the Fermi surface which connects these Weyl points near $E_F$ experimentally by ARPES. Our findings provide evidence for a TRS breaking induced Weyl state that is present in $YbMnBi_2$ but not $EuMnBi_2$.

In Fig.1 we present the crystal and electronic structures of both materials as well as some of their basic physical properties. $EuMnBi_2$ and $YbMnBi_2$ crystallize in I4/mmm and P4/nmm structures respectively which are very similar, but with an important difference in mutual arrangement of the basic structural units. The square Bi net (green atoms, Bi2) contains a mirror plane in the case of $EuMnBi_2$ and contains a glide plane in $YbMnBi_2$. Because of this difference the former has a body-centered and the latter a primitive tetragonal lattice and corresponding symmetry of the perturbing potential drastically alters the pristine dispersions of the states originating from the square Bi nets. In addition, the magnetic moment of Eu ions is considerably larger than that of the Yb ions.

The resistivities and magnetoresistances are compared in Fig. 1d,e. Both compounds are metallic and have large, non-classical magnetoresistances as seen for other $AMnBi_2$ compounds. In $EuMnBi_2$, a sharp feature in the magnetoresistance is seen due to a metamagnetic transition in the Eu sublattice, as reported in Ref. 20. For $EuMnBi_2$ the magnetism is very obvious as, at least below 300K the magnetism is only due to the Eu moments as the Mn moments are locked antiferromagnetically. Additional ordering of Eu atoms seen in resistivity at ~ 20 K implies that the ground state of $EuMnBi_2$ is pretty well defined and all magnetic moments, Mn and Eu, are antiferromagnetically ordered. There is no hysteresis in any of the measurements. In the $YbMnBi_2$ compound we see no strong magnetism between 50 and 300 K, however below 50K new features in the magnetization develop.

We show MH curves for $YbMnBi_2$ in Fig. 1f. The behavior above 20,000 Oe should be all bulk effects and the curves for 300K, 200K, 100K, and 50K only show a very weak either diamagnetic or weak paramagnetic response to the field above 20,000 Oe. This implies the spins are locked in possibly perfect AFM order. However, as we get to low temperature, the magnetic response picks up and becomes significantly more anisotropic signaling some new spin degrees of freedom. This would be a disruption of perfect antiferromagnetism due to e.g. canting and is an evidence for time-reversal symmetry breaking in YbMnBi$_2$.

We performed different types of band structure calculations. Corresponding Brillouin zones (BZ) are shown in the insets to Fig. 1a. In order to understand general aspects of the electronic structure of both materials, we first carried out calculations treating both Mn and rare earth originated states as quasicore (SI). When spin-orbit coupling (SOC) is neglected, this approach yielded the expected linear dispersions generated almost exclusively by Bi2 $p$ states, which cross near or at the Fermi level [22]. The momentum locus of such symmetry protected Dirac crossings is indicated inside the BZ as a green loop, a line and a point. Exactly such elements are needed to construct a Weyl state and both compounds indeed demonstrate a number of Dirac dispersions using this simple approach. Already at this stage it is clear that staggered geometry of the Yb atoms with respect to Bi net protects more Dirac crossings in YbMnBi$_2$ than the coincident geometry in EuMnBi$_2$.

The next step is to include the Mn $d$ magnetism and spin-orbit coupling to the computational scheme. As a first approximation, we consider the ordering of Mn moments as perfectly antiferromagnetic (Fig. 1a) and neglect the contributions of the 4$f$ states of rare earth ions which are still treated as non-spin-polarized quasi-core states. The results of the fully relativistic band structure calculations are presented in Fig. 1b. Three bands contribute to the formation of the Fermi surface (FS), 2D cross-sections of which are shown in Fig. 1c. Green and blue dispersions correspond mostly to Mn states while the red one represents discussed above Bi $p$-states. As is seen, spin-orbit and exchange interactions make the Dirac electronic states "massive" in both materials, i.e. opens the gaps in most parts of the BZ. Again, there are important differences. While the gap is equally large (~ 0.2 eV) across the BZ in EuMnBi$_2$ creating isolated "lenses" of FS, the gap in YbMnBi$_2$ is smaller (down to ~10 meV along the ΓM line), adding to the "lenses" small electron-like pockets near the X-points. The distinct separation of these two portions of the FS signals the opening of the gap also in YbMnBi$_2$ and, in particular, breaking of the nodal loop (green contour in upper BZ sketch in Fig.1), but above $E_F$ one crossing is preserved as is seen from Fig. 1b where the red and blue dispersions touch each other along ΓM direction. We have found in total four 3D Dirac points in YbMnBi$_2$ with the coordinates (±0.193,± 0.193, 0.015) and (±0.193,-+ 0.193, -0.015) which are schematically shown in lower BZ sketch in Fig. 1 as green points.

In Fig.2 we present the results of comparative study of Fermi surfaces and underlying dispersions by angle-resolved photoemission spectroscopy (ARPES). The FS of EuMnBi$_2$ (Fig. 2a) is remarkably similar to that resulted from the simplified calculations shown in Fig. 1c, apart from that Mn-states are absent at the Fermi level. This is not surprising since their computed energy localization is very sensitive to the value of the U parameter usually used to correctly reproduce the correlated nature of Mn 3$d$ states. The well separated four lenses with sharp corners are clearly observed and depend only weakly on photon energy (SI) confirming their origin from Bi $p$ states from the two-dimensional square net. The ARPES intensity map taken along the cut #1 illustrates an ideally linear on the scale of 1 eV behavior of these features with enormous Fermi velocity (9 eVÅ). Other states with considerably larger $k_z$ dispersions (which make them appear blurred because of intrinsically moderate $k_z$ resolution) are also seen, in accordance with the calculations. The cuts #2 and #3 emphasize another agreement with the theory --- Dirac crossings are destroyed at all occupied momenta and sizeable energy gaps are opened in the vicinity of the Fermi level.

A qualitatively different picture is observed in the case of YbMnBi$_2$ (Fig. 2b). Although bearing a certain degree of similarity with the simplified calculations, the Fermi surface looks continuous and has a fine structure. The intensity distribution recorded along the cut #1 also shows extremely fast and sharp in momentum states originated from the Bi-net, as in EuMnBi$_2$. Cut #2 runs through the additional FS elements and demonstrates the crucial dichotomy with the material containing Eu: the Dirac crossings appear to remain protected or the gap is much smaller. At particular photon energies (20.5, 27, 36 or 55eV, see SI) the experimental FS contour of YbMnBi$_2$ looks sharp and continuous, i.e. hole-like "lenses" are connected to small electron-like pockets near X-points, implying the survival of the Dirac crossings in the momentum regions where these two FS sheets appear to be separated in Fig. 1c, at least for particular $k_z$. Here the crossings coincide with the Fermi level and large gaps would be detectable. The continuous shape of the Fermi surface Fig. 2b is the consequence of the protection of the Dirac crossings at least in some points along the Fermi surface contour. By symmetry considerations, these Dirac crossings should not exist in the presence of SOC and TRS and inversion symmetries. Given no evidence of structural transitions, it is likely that TRS is broken in YbMnBi$_2$ also purely from the ARPES measurements of FS.

Obviously, it is instructive to take the data along the cut which contains these special momentum regions. The next panel of Fig. 2b (cut #3) represents a crucial dataset which underlines the exotic electronic structure of YbMnBi$_2$ and directly proves the presence of the doubly (not quarterly) degenerate Dirac points, i.e. the long sought Weyl points. The first striking observation is that the Bi dispersions appear to be non-spin-degenerate in this particular region of the BZ. This breaking of the spin-degeneracy is key, and is not consistent with the AFM calculations. Indeed, the dispersion is seen to split between 100-200 meV binding energy and only one pair of split features is gapped while the other still reaches the Fermi level where they clearly cross. It is these crossing points sitting at the Fermi level are the essential components of the electronic structure of a Weyl semimetal. Since these points are also the places where the electron-like FS touches the hole-like FS, YbMnBi2 is a canonical example of type II Weyl semimetal [21]. We note that we managed to observe such a clear picture of crossing of non-degenerate bands only at particular photon energies (20, 27 and 31 eV, see SI).

In order to understand the magnetism seen in the MH curves and the origin of the observed lifting of the degeneracy by ARPES (TRS breaking), we have carried out the calculations considering several types of canted antiferromagnetism of Mn atoms - with the in-plane and out-of-plane ferromagnetic components. Remarkably, in all cases the degeneracy was lifted in the way it is observed experimentally: the gap between the linear Dirac dispersions became smaller for one of the non-degenerate components and disappeared in several momentum points. In other words, canting resulted in splitting of the 3D Dirac points (lower left inset to Fig. 1) into Weyl points and Weyl loops. In Fig. 2c we present the main results for the configuration of spins with the ferromagnetic component along ΓM direction (ΓM$_\text{II}$). Two sets of Weyl points are observed in this case. The first one consists of four points with the coordinates (0.193,0.193,0.12), (-0.193,-0.193,0.12) and (0.194,0.194,-0.09), (-0.194,-0.194,-0.09) with the energy of 168 meV above the Fermi level. These points move towards the mentioned earlier 3D-Dirac points upon decreasing of the canting angle and annihilate in the case of perfect antiferromagnetism (see SI and Fig. 2c). Another set of Weyl points with coordinates (0.394,0.045,0.131), (0.394,-0.045,0.131), (0.045, 0.394,0.131) and (-0.045, -0.394,0.131) with the energy of 40 meV below the Fermi level corresponds to the crossing of the non-degenerate bands observed experimentally in Fig. 2b. In the right panels of Fig. 2c we show the band dispersions in the vicinity of one of these points. The key observation is that the Weyl "cones" are very anisotropic. If perpendicular to the Fermi surface in $k_xk_y$ plane, the Fermi velocity is a record one (9 eVÅ), if along the Fermi surface or along $k_z$, it is more than two orders of magnitude smaller (~0.043eVÅ). Apart from the mentioned Weyl points we found the loops of Weyl crossings around the $\Gamma M_\perp$ (these are analogous to the Dirac loop shown earlier in Fig. 1, but doubly degenerate, not quarterly, shown in Fig. 2c) and very small gap regions above the second set of Weyl points close to the ZAR plane. Our band structure calculations thus clearly point to the first realization of the Weyl state in a canted antiferromagnet (SI).

Taking into account the importance of the $k_z$-dispersion and very small gaps caused by anisotropic Weyl cones implied by the calculations we have performed both very detailed photon energy dependent measurements and high-resolution measurements near the calculated Weyl points (Figs. S10 and S15 in SI). It turned out that the mentioned earlier photon energies at which the FS contours were sharp and continuous approximately correspond to $k_z$-locations of either second set of Weyl points or the minimum gaps, naturally explaining the observed effect since the $k_z$-dispersion in these regions is negligible. However, even at hν=20 eV we could not clearly resolve the tiny gaps and in both cases we observed qualitatively similar distribution of spectral weight near the Fermi level, as if the Weyl points existed at both $k_z$s.

As is shown in Fig. 2c, the Weyl points from the first set are projected to the surface BZ (dashed circles) such that they annihilate. The Weyl points from the second set, in contrast, should be connected by peculiar surface states, the hallmark of the Weyl state, when measuring the 001 surface, as in our experiment. In order to detect the possible fine structure and elusive arc of surface states, we have used lower photon energies to increase momentum resolution. The high-precision Fermi-surface map is shown in Fig. 3a. In the lower part of the map we overlay the experimental intensity with the guides to eye to explain the observed features. There are clearly more features than one could previously conclude from Fig. 2b and simplified calculations presented in Fig. 1. While one can still identify the elements of the calculated FS – hole-like lenses and electron-like pockets close to the X-points, they appear connected and in the region of the "lens" one can clearly identify three features instead of two. If the connection between the lenses and electron-like pockets, as was shown earlier, is due to the lifting of the spin-degeneracy, the third feature on the map is the Fermi surface arc which connects two previously detected Weyl points. The data shown in Fig. 3 b,c confirm this assignment. Indeed, there are three Dirac-like dispersions corresponding to each "lens". These two datasets are taken using different photon energy in an attempt to distinguish the bulk from the surface component. One can notice that in panel b two internal lines are less intense than the external ones and in panel c the most inner ones are weaker than four others. On the basis of this observation and analysis of many other similar datasets we arrived at the conclusion that it is innermost linear dispersion corresponds to the Weyl arc. This is supported by the data taken from $EuMnBi_2$. We present enlarged fragments of panels b and c together with similar cuts taken from different sample in Fig. 3d. The third feature is clearly present in all datasets. In Fig. 3e the high-precision Fermi surface map is zoomed in to the single lens. Not only is the arc remarkably absent, but also the shape and intensity distribution of the ARPES intensity helps us to identify the bulk component in $YbMnBi_2$. The bulk lens has a very strong straight section and corners. Exactly the same characteristics are peculiar to the outer dispersions in the discussed earlier panels (a) and (b) of Fig. 3. As expected, the energy-momentum cut in Fig. 3f contains only four dispersions, i.e. no signature of Weyl arc.

We briefly discuss other possible explanations of the unusual behavior seen in $YbMnBi_2$. We would like to immediately exclude the possible presence of slightly disoriented crystallites. First, the single crystals are of exceptional quality and at no stage during their characterization have we noticed the presence of such crystallites. Second, in this case one would observe doubling of all the features and we see only three. We also believe that the combined influence of finite $k_z$-resolution and $k_z$-dispersion cannot be responsible for the observed effect (the z- axis is perpendicular to the surface). The width of the linear dispersing features is resolution limited for all used photon energies (Fig. S14 in SI) and $k_z$-dispersion related artifacts are usually much broader and not that clearly separated. In addition, the $k_z$-dispersion is very similar in Eu-containing material (see Fig. 1 b) and there we do not see any evidence for additional features. Finally, we rule out the "technical" influence of the surface, such as relaxation [23], since, again, EuMnBi$_2$ cleaves in similar way and there one would expect more surface-related problems because of much larger magnetic moments. We stress here that the decisive experimental fact, which speaks in favor of our interpretation, is the number of features at the Fermi level: any kind of artifact, including the relaxed surface, would produce a replica of the "lens" and thus double the number of crossings. We observe a "lens" and an arc, and the dispersion corresponding to the flat part of the former is a single feature at all photon energies (Fig. S14 in SI). Previous ARPES study on closely related materials SrMnBi$_2$ and CaMnBi$_2$ isostructural to EuMnBi$_2$ and YbMnBi$_2$, respectively, did not reveal any surface states, but strongly anisotropic Dirac dispersions.

We make now an attempt to illustrate the origin of the revealed Weyl state in YbMnBi$_2$ in Fig. 4 considering the gradual modification of the electronic and crystal structure of square net of Bi atoms. We start from the purely two-dimensional construct – the Bi net itself. Its electronic structure is shown in Fig. 4a. In this larger BZ there are no Dirac crossings and the dispersions are simple solutions of the tight-binding approach which takes into account only Bi *p* states. Doubling of the unit cell because of the staggered coordination by Yb atoms results in Dirac crossings [22] exactly at the Fermi level and the momentum locus of these points is given by the diamond-shaped Fermi contour. This unique electronic structure is shown in three dimensional space in the lowest panel of Fig. 4b and it is characterized by the 8 times degenerate electronic states at X-point. It is remarkable that it is in the proximity of this point the degeneracy will be lifted completely in YbMnBi$_2$. In the next column of panels (Fig. 4c) we show the case of Eu material with SOC and perfect AFM included. This example shows why Weyl state cannot be realized in EuMnBi$_2$: spin-orbit and exchange interactions open the energy gaps at all Dirac points. Here it is also easy to see the mechanism of the formation of the lenses: the gap is not constant and is not centered at the Fermi level. Next set of panels (Fig. 4d) represents the case of YbMnBi$_2$ as it is seen from ARPES. The value of $k_z$ corresponds to the plane containing Weyl points and the constant gap centered at the Fermi level is considered for simplicity. As mentioned earlier, the degeneracy is seen to be lifted leading to formation of four pairs of Weyl points. Note that the cut ΓX''runs through the Weyl point. If the gap would be centered at the Fermi level along the diamond-like contour (Fig. 4b), YbMnBi$_2$ would have been a canonical Weyl semimetal of type I. Presence of lenses and electron pockets near X points which touch in a single point make it the Weyl semimetal of type II. Finally, we present a schematic plot of the electronic states responsible for the Fermi arc together with the bulk-originated states in the panels of Fig. 4e. Again, for simplicity, we do not show all bands and the gap is chosen to be constant and centered at $E_F$. The arc at the Fermi surface is given by the surface states shown by green color. They originate from the bulk states forming the one Weyl point and terminate in bulk states in the vicinity of another Weyl point.

It is interesting to see that arc connects not the closest Weyl points on the experimental maps. However, the experimental picture is most likely a superposition of two domains and the connection is indeed between two closest Weyl points as in the calculations (Fig. 2c). Moreover, the slow dispersion along the Fermi surface contours naturally explains the large separation of the Weyl points – even insignificant lifting of the degeneracy would split 3D Dirac points in widely separated Weyl points.

Having demonstrated the realization of the Weyl state in YbMnBi$_2$ experimentally and theoretically, we would like to emphasize that further studies of this material are called for. It is not clear which exactly magnetism results in time-reversal symmetry breaking and how to handle it theoretically. Our calculations demonstrate that canted antiferromagnetism is one of the possible solutions, but the exact configuration of spins is to be found. The chirality of Dirac electrons has already been discussed in related material with Sr instead of Yb (Ref. 22). Due to extremely difficult ARPES experiments (low photoemission signal) and very high resolution needed to resolve the arcs, the use of spin-resolved modification of the technique is nearly blocked, meaning that direct observation of the chirality of the Weyl points is probably postponed until the further improvement of resolution of spin-resolved ARPES. However, the presence of the time reversal symmetry breaking below 50 K, the existence of a continuous Fermi surface, a two-fold degenerate crossings at the Fermi level, and Fermi arc surface states that connect these crossings across the surface BZ are direct evidence of a TRS breaking induced Weyl state in YbMnBi$_2$ strongly supported by the band-structure calculations of the canted antiferromagnetic state; future work would be helpful in elucidating the exact mechanism of creating this state in this material.

*Acknowledgements*

We are grateful to Denis Vyalikh, Geunsik Lee for the fruitful discussions and to Alexander Fedorov, Yevhen Kushnirenko and Erik Haubold for the help at the beamline. The research at Princeton was supported by the ARO MURI on topological insulators, grant number W911NF-12-0461. We acknowledge Diamond Light Source for time on I05 under proposal SI11643-1.

*Experimental and calculations*

High-quality single crystals were grown using a Bi rich melt in the ratios YbMnBi10 and EuMnBi10. The elements were heated to 1000C and cooled to 400C at 0.1C/min, then subsequently centrifuged to remove excess Bi. The crystals exposed mirror-like portions of the surface after the cleave in ultra-high vacuum breaking z-periodicity.

The structures of the compounds were solved by single crystal x-ray diffraction (Table S3).

ARPES measurements were performed at the I05 beamline of Diamond Light Source, UK. Single crystal samples were cleaved in-situ at a pressure lower than $2*10^{-10}$ mbar and measured at temperatures about 7K. Measurements were performed using (s,p)-polarised synchrotron light from 18-100 eV and employing Scienta R4000 hemispherical electron energy analyser with an angular resolution of 0.2-0.5° and an energy resolution of 3 - 20 meV.

Band structure calculations were performed for experimental crystal structures of YbMnBi$_2$ and EuMnBi$_2$ using the relativistic linear muffin-tin orbital method as implemented in the PY LMTO computer code [24]. Perdew-Wang [25] parameterization of the exchange-correlation potential in the local density approximation (LDA) was used. Localized 4f states of a rare earth ion were treated as quasi-core states assuming Yb2+ (f14) and Eu2+ (4f7) configurations.

AFM calculations were performed assuming that two Mn ions in the unit cell have opposite magnetization directions, i.e., without increasing the structural unit cell. For both compounds this results in checkerboard AFM order in ab plane. Magnetic order along c axis is, however, FM for YbMnB$_2$ (P4/nmm) structure and AFM for EuMnB$_2$ (I4/mmm).

**Figure 1. Calculated electronic structure and basic properties of $EuMnBi_2$ and $YbMnBi_2$.** a) Crystal structures of both materials together with the corresponding Brillouin Zones. Green lines and a point are a locus of Dirac crossings according to the calculations not taking into account spin-orbit interaction (SM). Red arrows are magnetic moments of Mn atoms. Inset shows the locations of 3D Dirac points found in the AFM-SOC calculations. b) Results of the fully relativistic band structure calculations taking into account perfectly antiferromagnetic arrangement of Mn moments. c) Corresponding Fermi contours for ΓXM and TNP high symmetry planes. d) Resistivities and e) Magnetoresistances of $EuMnBi_2$ and $YbMnBi_2$. f) Magnetization curves showing the onset of the magnetism which breaks time-reversal symmetry in $YbMnBi_2$.

**Figure 2. Experimental electronic structure of $EuMnBi_2$ and $YbMnBi_2$.** a) Fermi surface map of $EuMnBi_2$. White arrows with numbers indicate the directions along which the spectra in the panels to the right have been recorded. Data are taken using 27 eV photons (cut #1 at 44 eV). No Dirac crossings are seen in this case. b) Same as a), but for $YbMnBi_2$. Zero momentum in panel with the cut #2 (hν = 80 eV) corresponds to the BZ center. Dashed lines in the panel with cut #3 are guides to eye representing splitting of the degenerate bands due to TRS breaking. Crossings of these bands near

the Fermi level are Weyl points. c) Schematic results of the band-structure calculations taking into account the canting (10°) resulting in in-plane ferromagnetic component along the (1,1,0) direction. $k_{x*}$ and $k_{y*}$ are the directions perpendicular and along the Fermi surface contour respectively.

**Figure 3. Fermi surface arc.** a) High precision Fermi surface map of YbMnBi$_2$ at hν = 20.5 eV showing unusual number of features (brown lines) which can be explained by the presence of the arc (dotted line) corresponding to the surface states connecting the Weyl points (green points). b) Spectra corresponding to cut #1 shown in panel a) by white arrow (hν = 20.5 eV). White curve is the momentum-distribution curve at the Fermi level showing the presence of six crossings. The inner ones are due to the arc. c) Same as in b) but for hν = 27 eV. d) Zoomed in fragments of panels b) and c) as well as similar data for other photon energies. e) High-precision Fermi surface map of EuMnBi$_2$. f) Spectra along the cut #2 indicated by white arrow in e). Photon energy is 27 eV. No sign of the arc is present in the data.

**Figure 4. The way to Weyl.** Schematic electronic structures corresponding to the structural fragment shown in the upper panel (from top to bottom): Fermi contour, band structure along selected directions, 3D E(**k**) representation in the whole BZ. a) 2D network of Bi atoms. b) Bi-network surrounded by Yb atoms in staggered geometry. c) Bi-network surrounded by Eu atoms in coincident geometry. Spin-orbit coupling and perfect antiferromagnetism are taken into account in the calculations. White dashed lines in the lowest panel indicate where the sheets cross the Fermi level. d) Case with Yb and canted AFM which results in lifted degeneracy and creation of Weyl points. Inset to the lowest panel shows the cut through one of the Weyl points. e) Presence of the surface induces the non-trivial surface states (green planes) connecting the Weyl points (green points).

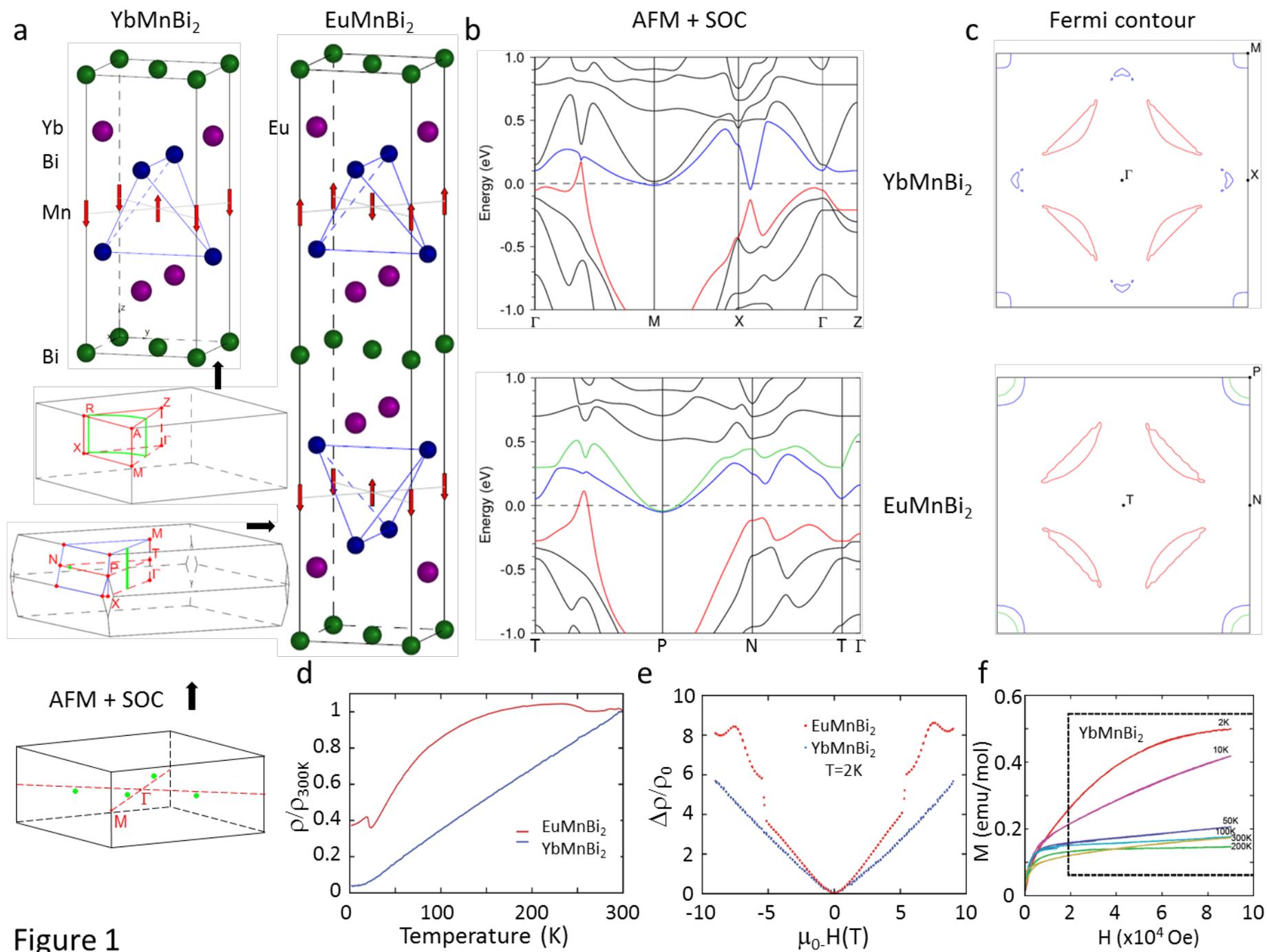

Figure 1

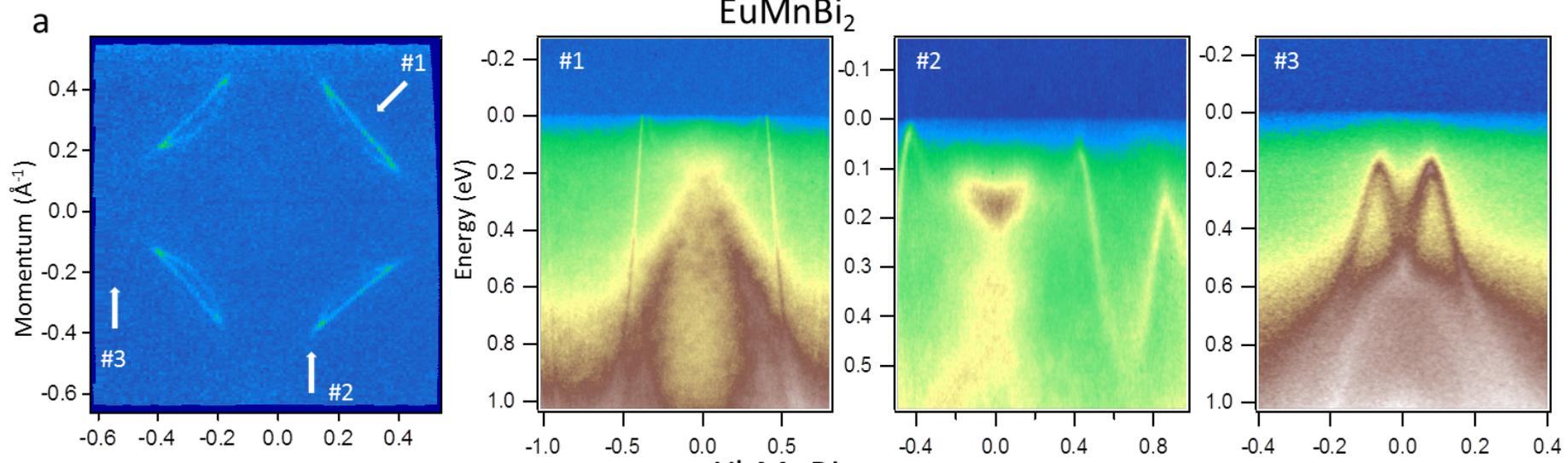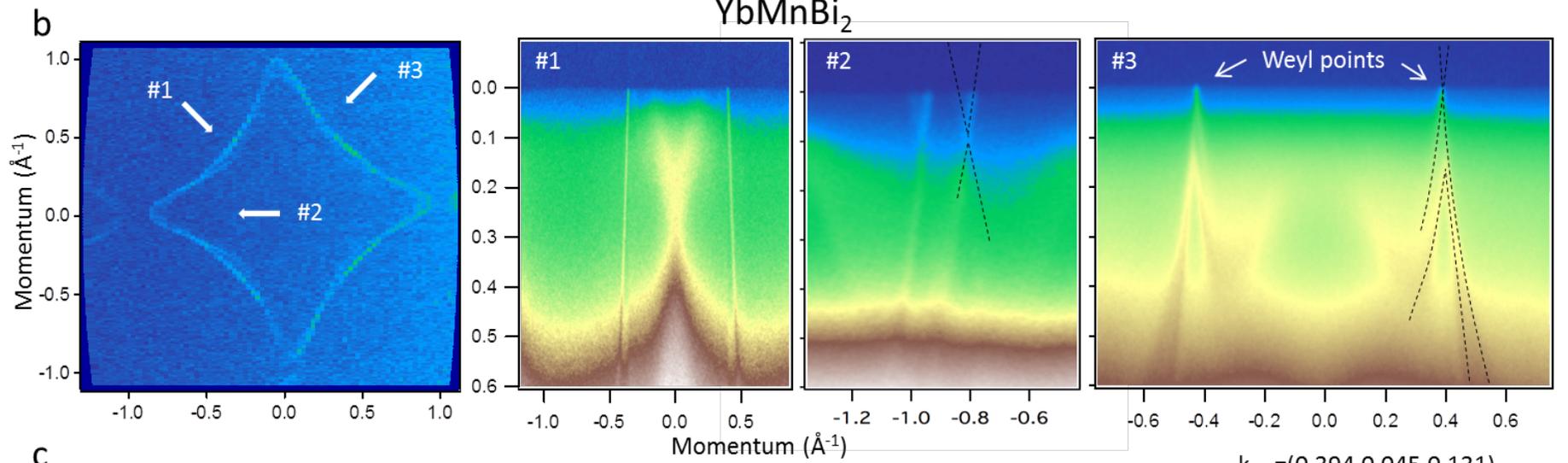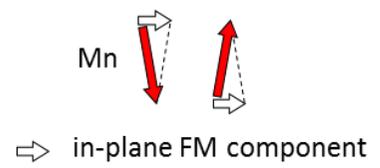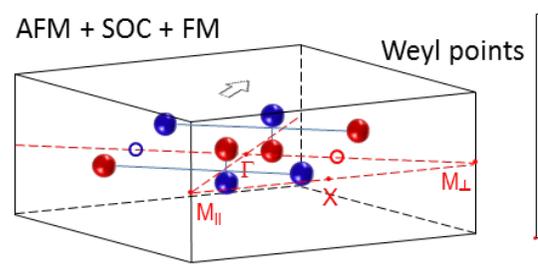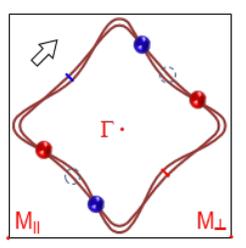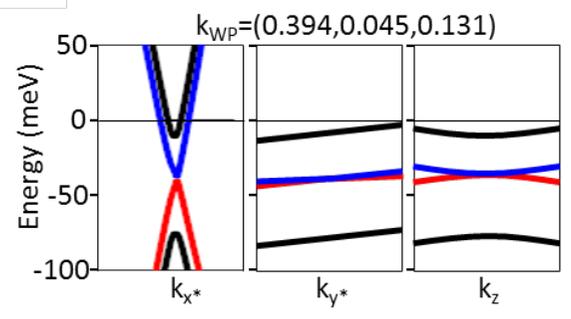

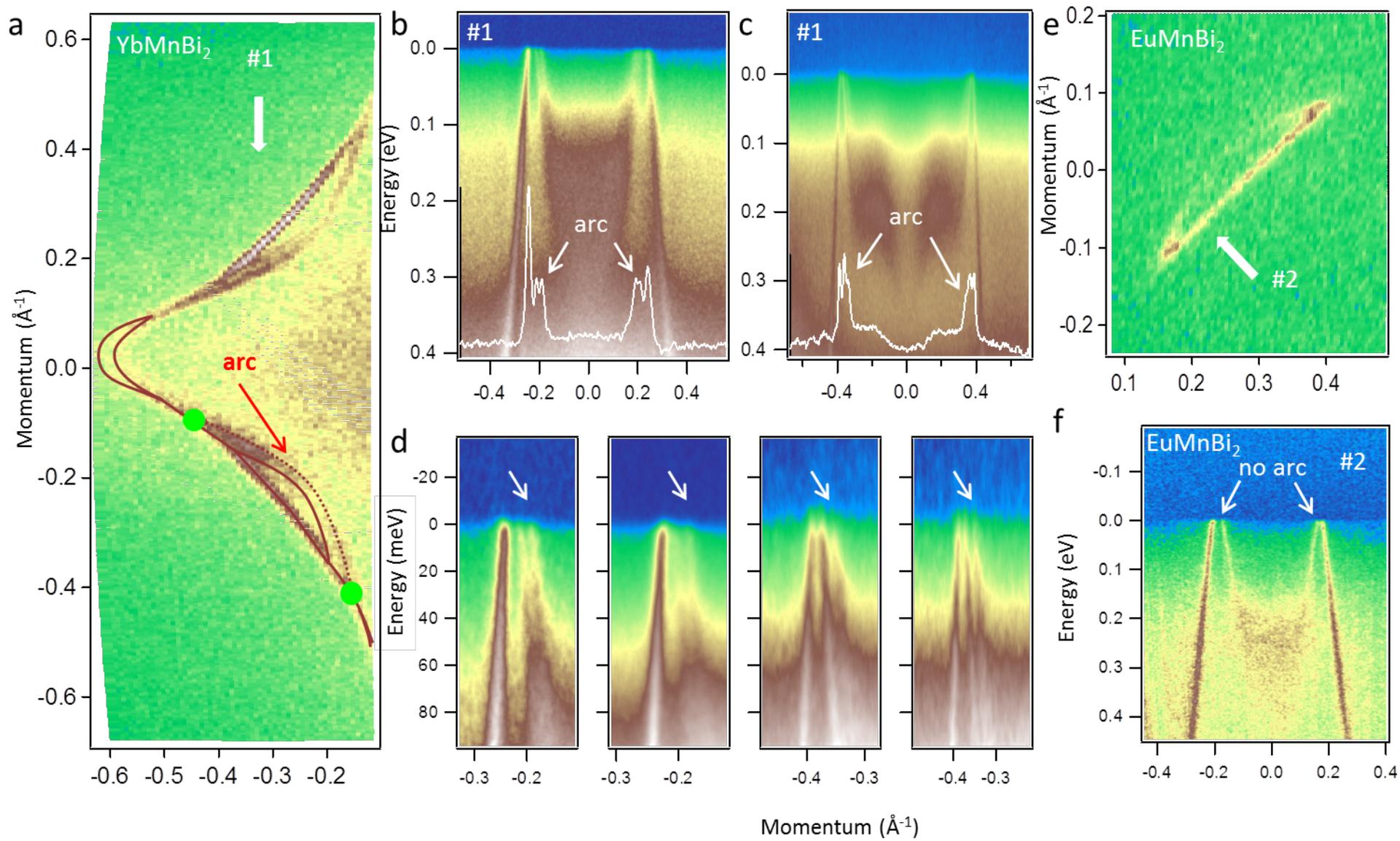

Figure 3

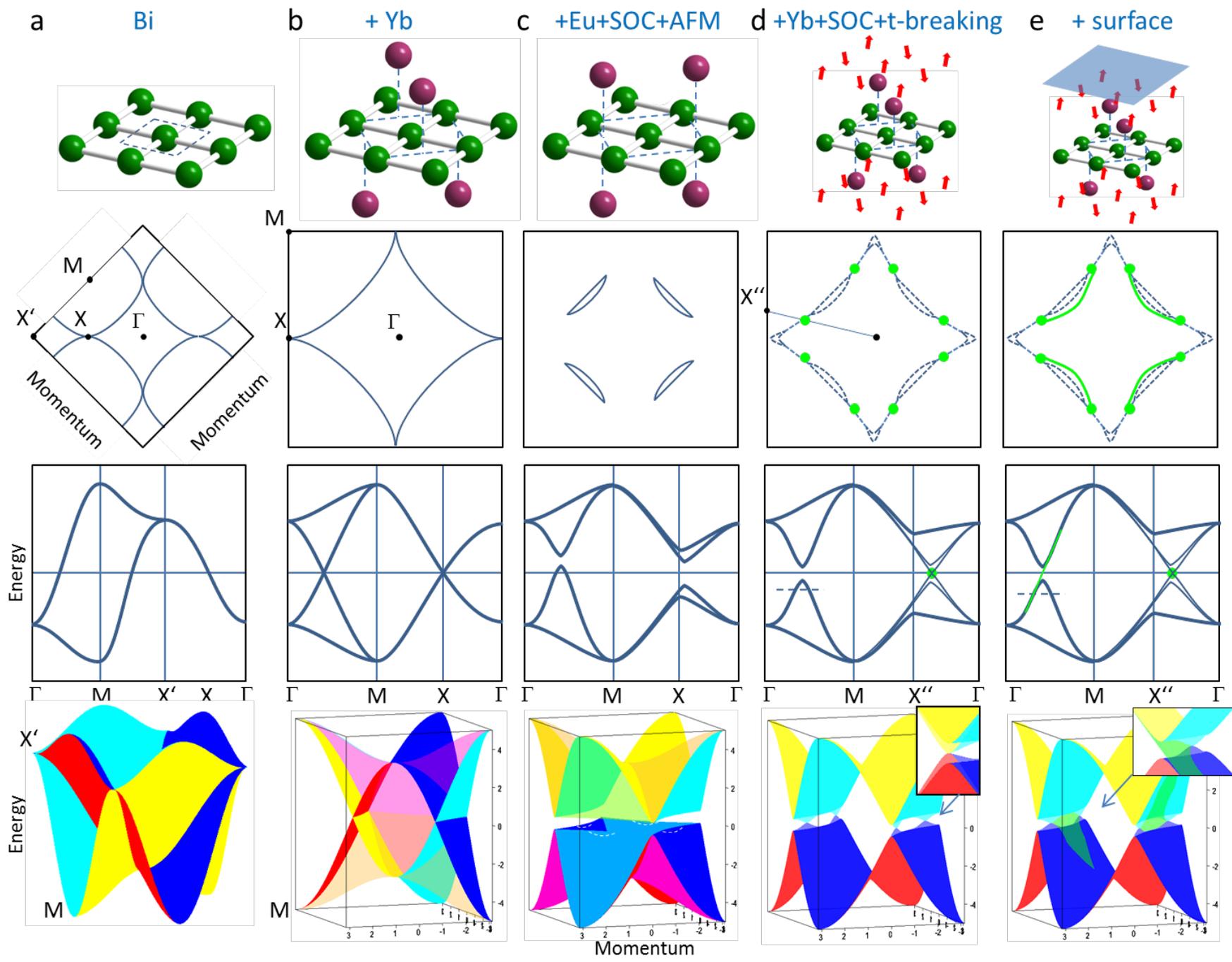

Figure 4

# Supplementary information

In this section we present additional datasets which support the material presented in the main part and help to understand the details of calculations and experiments.

THEORY

In Fig. S1 we present the result of the calculations without SOC and spin-polarization and treating Yb $4f^{14}$ and Mn $3d^5$ states as quasi-core states.

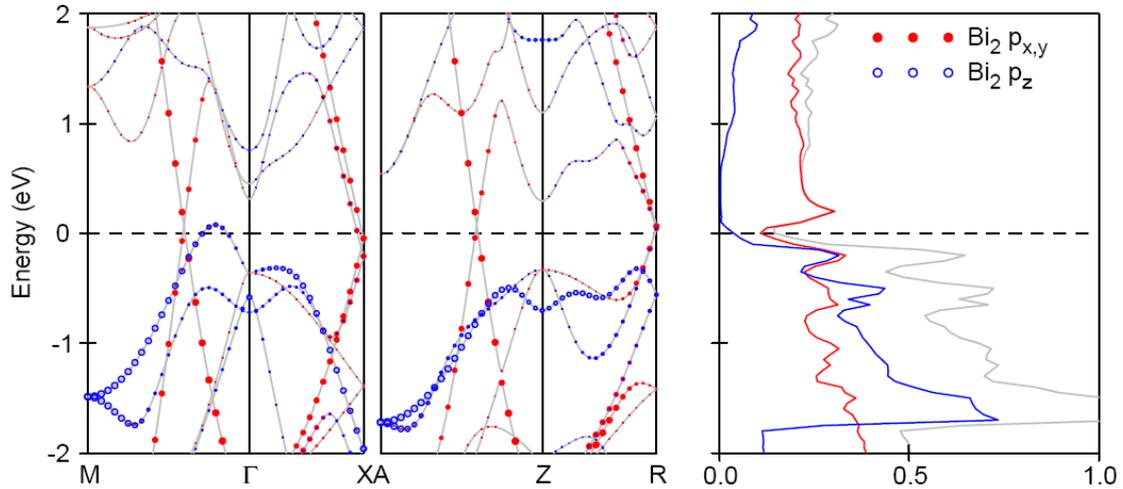

**Figure S1.** Calculations of YbMnBi$_2$ without SOC, Yb 4f and Mn d states. Bi$_2$ are the bismuth atoms from the 2D network.

As expected, bands at EF are formed by bismuth 2 p states from the two-dimensional networks. There are clear Dirac-like crossings along Γ(Z)–M(A) and Γ (Z)–X(R) near X. In order to find the location of all such crossing throughout the whole BZ, we have done the calculations along many directions and show the result in Fig. S2.

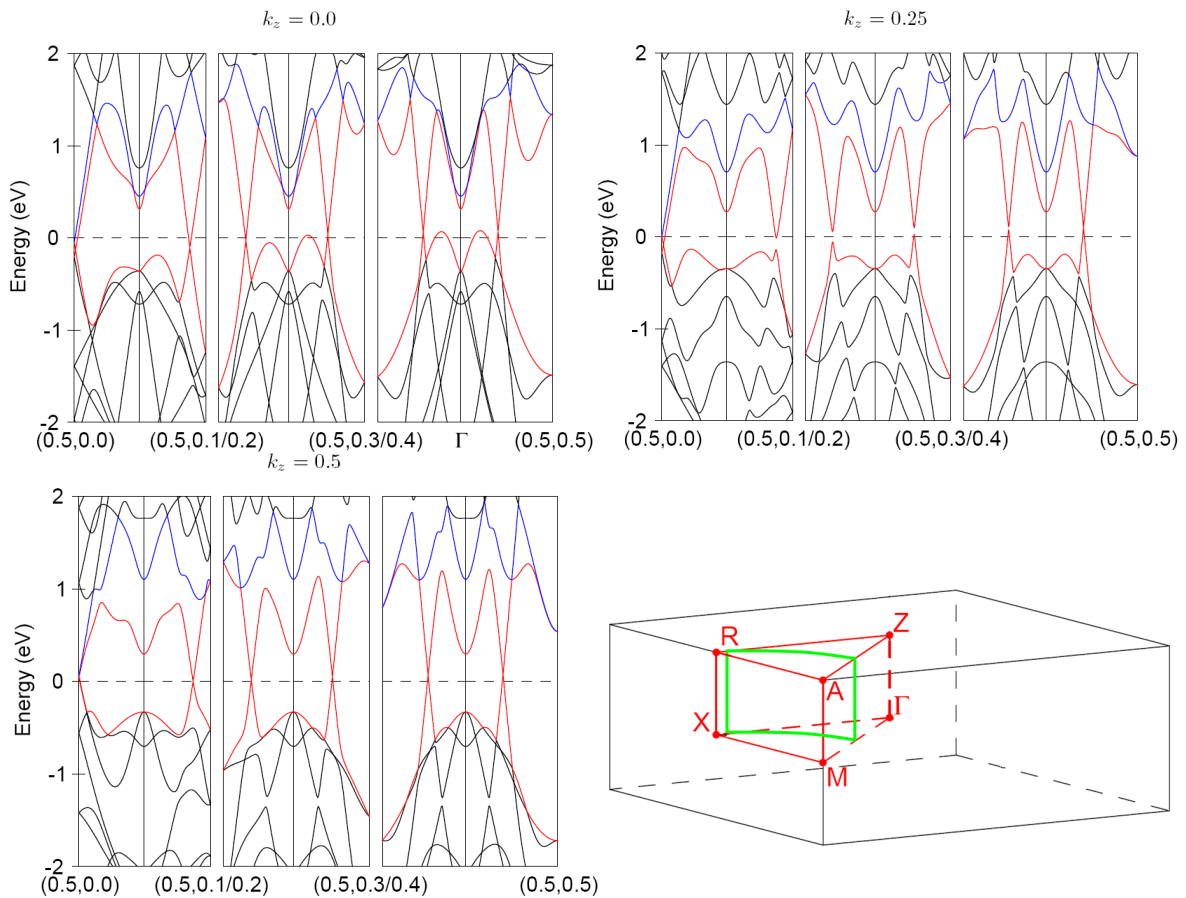

**Figure S2.** Bands of YbMnBi$_2$ calculated along (0,0,k$_z$)–(0.5,k$_y$,k$_z$) lines with k$_y$ = 0,0.1, . . . 0.5, and k$_z$ = 0, 0.25, 0.5. Green loop in the irreducible part of the BZ shows the locus of Dirac-like crossings.

The crossings along (0,0,k$_z$)–(0.5,0,k$_z$) and (0,0,k$_z$)–(0.5,0.5,k$_z$) lines survive for arbitrary k$_z$. Thus, they should be protected by vertical glide mirror planes M$_x$ and M$_{xy}$, respectively. Crossings for k$_z$=0, 0.5 and for an arbitrary k$_y$ should be protected by a horizontal glide mirror plane M$_z$.

Similar calculations for EuMnBi$_2$ are shown in Figs. S3 and S4.

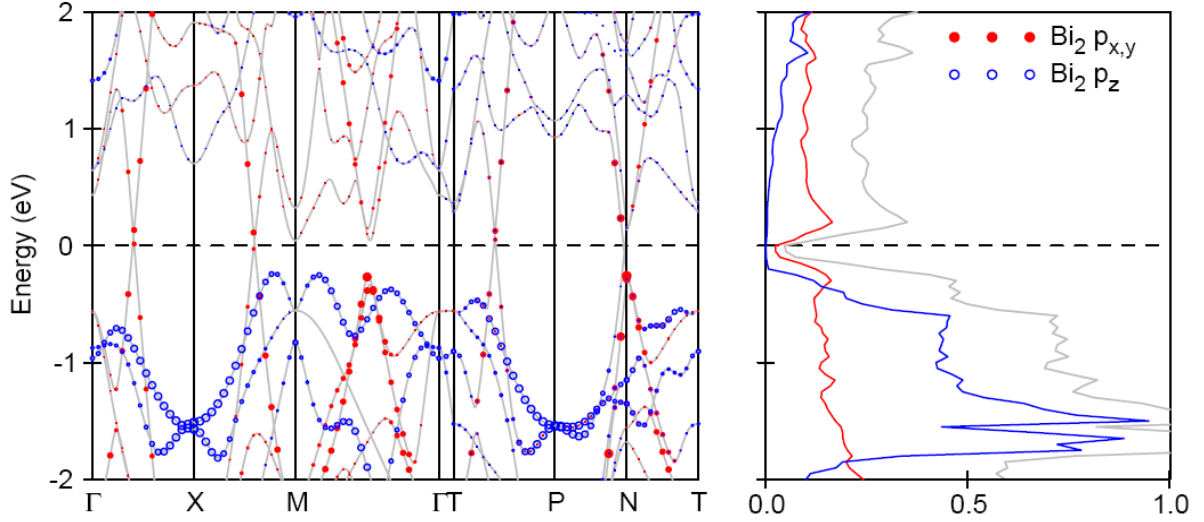

**Figure S3.** Calculations of EuMnBi$_2$ without SOC, Eu 4f and Mn d states. Bi$_2$ are the bismuth atoms from the 2D network.

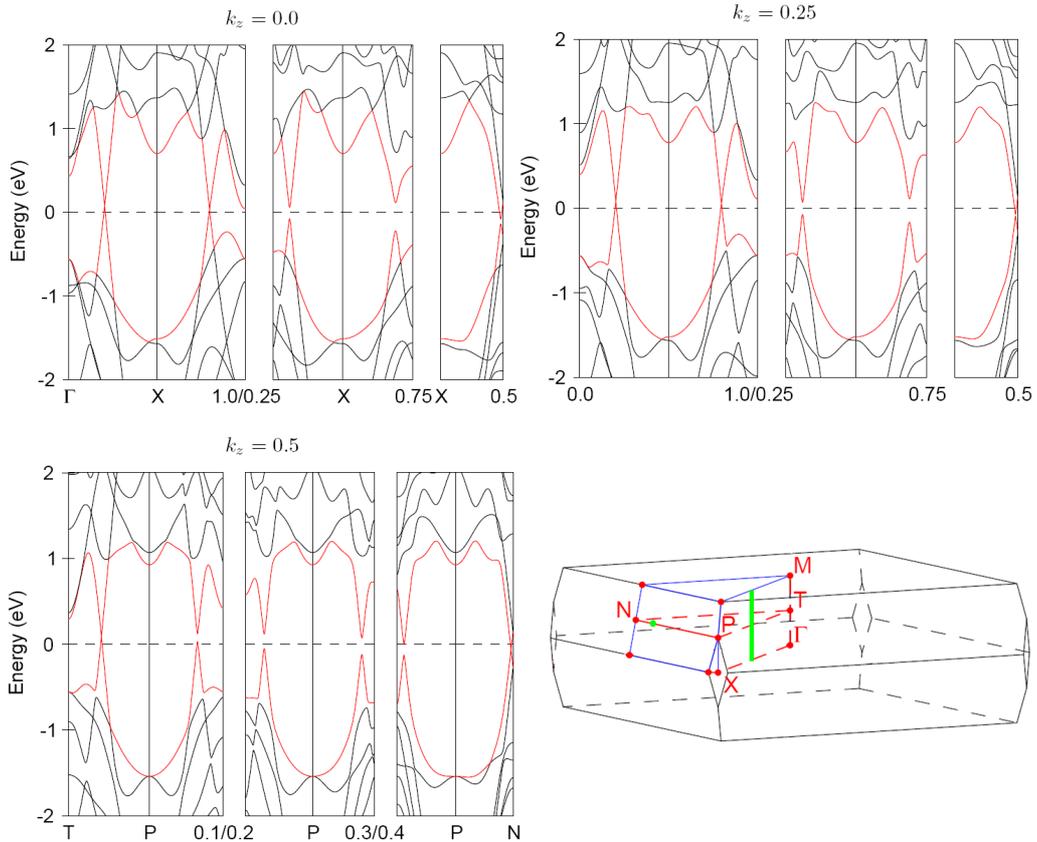

**Figure S4.** Bands of EuMnBi$_2$ calculated along (0.5,0.5,k$_z$)–(k$_x$,0,k$_z$) lines with k$_x$ = 0,0.25,0.5,0.75 and k$_z$ = 0, 0.25 and along P – (k$_x$,0,0.5) with k$_x$ = 0,0.1,0.2,0.3,0.4,0.5. Green line shows the locus of Dirac-like crossings. Green point is a 3D Dirac point.

In this case linear dispersions cross only in a vertical Γ-M–P–X plane and along N–P line creating a nodal line and 3D Dirac point near N-point.

Upon inclusion of SOC and AFM into the computational scheme all Dirac crossings become gapped in EuMnBi$_2$. In contrast, in YbMnBi$_2$ four 3D-Dirac points are observed. We show the band dispersions in the vicinity of one of them which lies in ΓMAZ plane in Fig. S5.

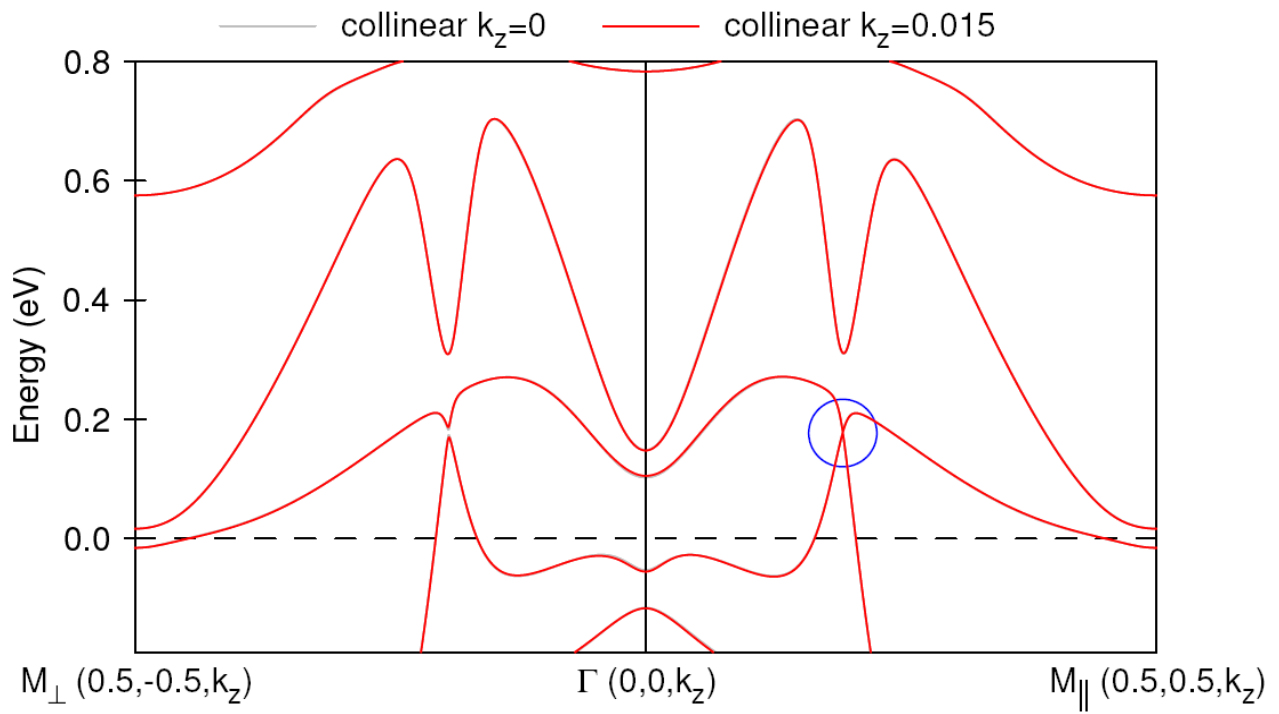

**Figure S5.** 3D-Dirac point in YbMnBi2.

Other symmetry related 3D-Dirac points are located at (-0.193,-0.193,0.015), (0.193,-0.193,-0.015) and (-0.193, 0.193, -0.015). All tthese points are schematically shown in the inset to Fig.1.

In order to explain the splitting of the bulk bands in YbMnBi2 seen experimentally we have carried out the calculations with canting. Canting angle is 10°. It produces net FM magnetization along (1, 1, 0). The results are shown in Fig. S6.

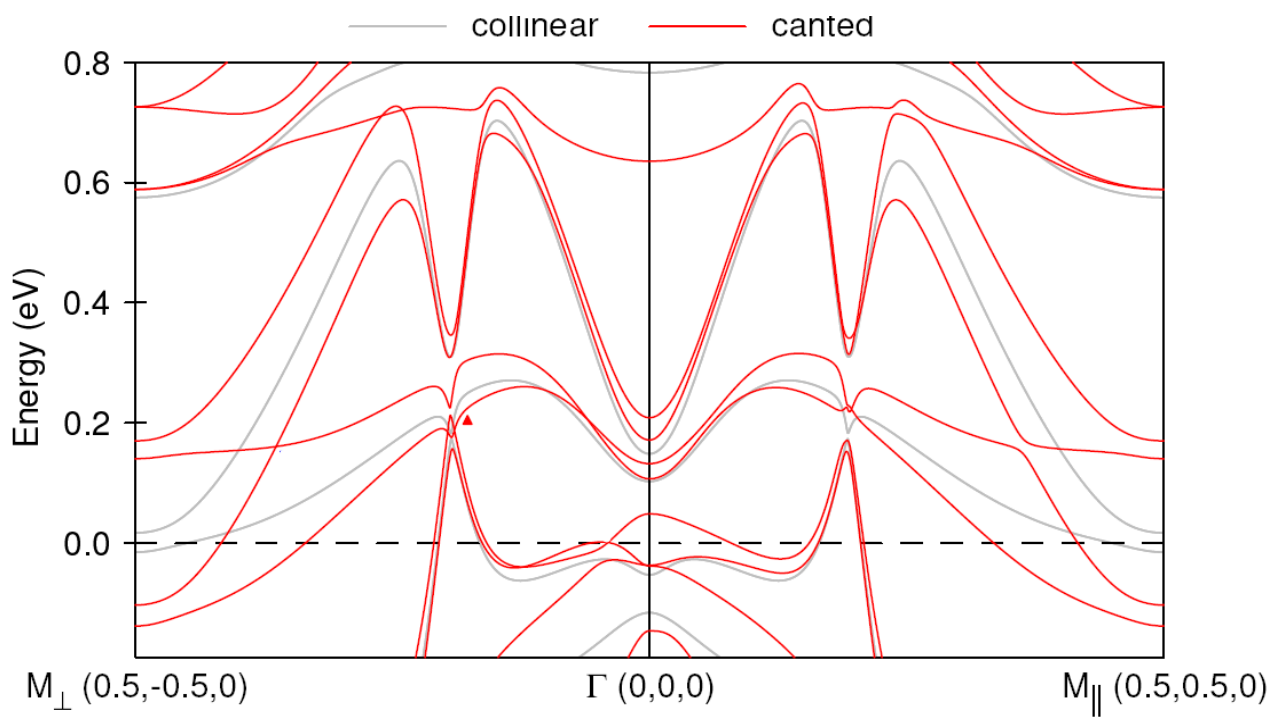

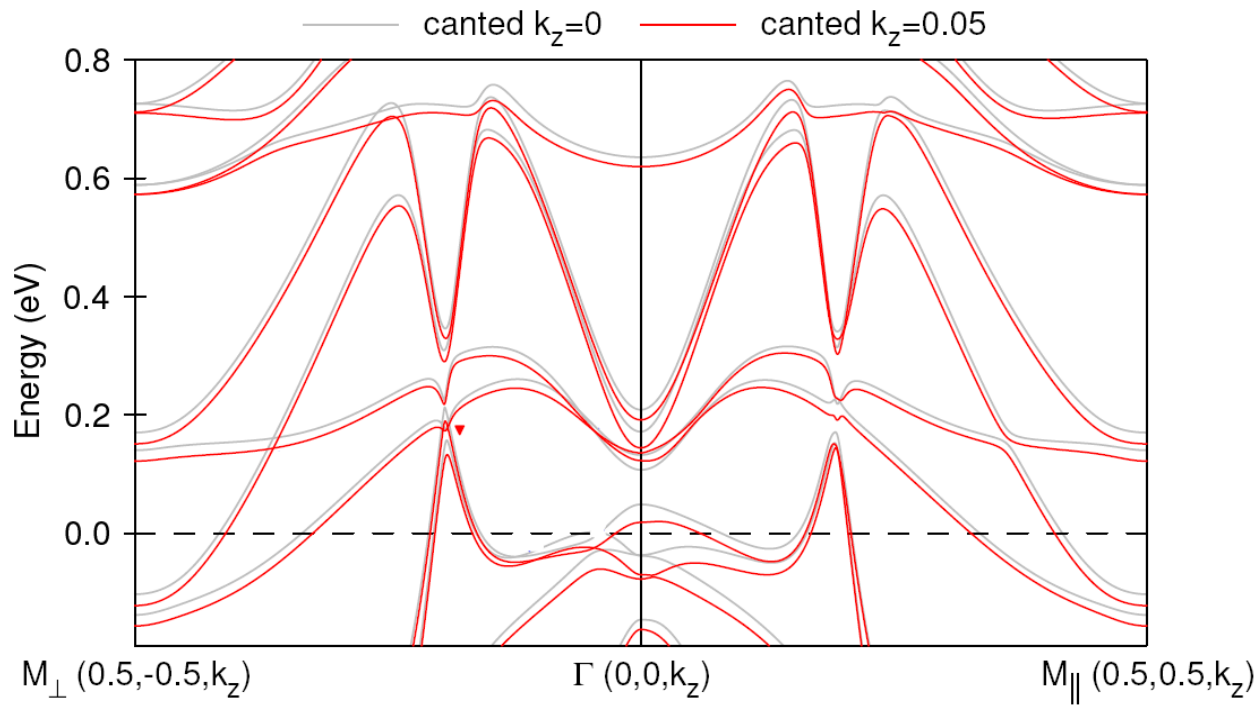
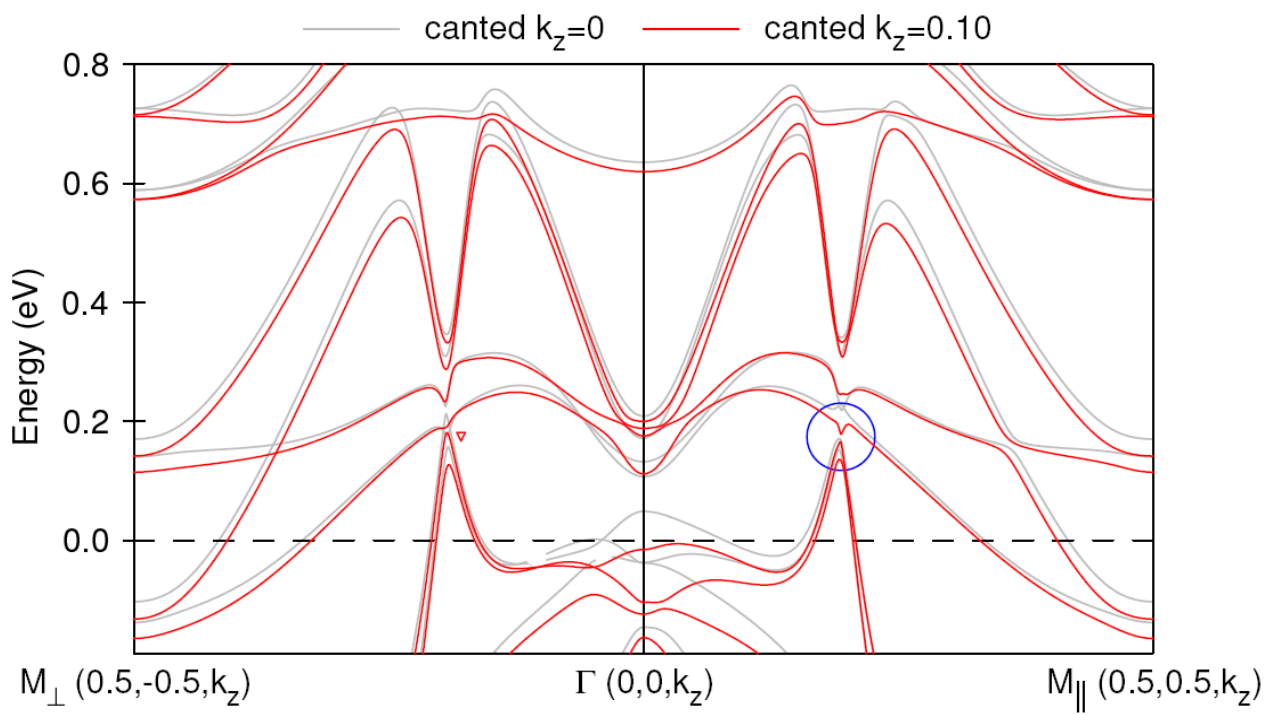

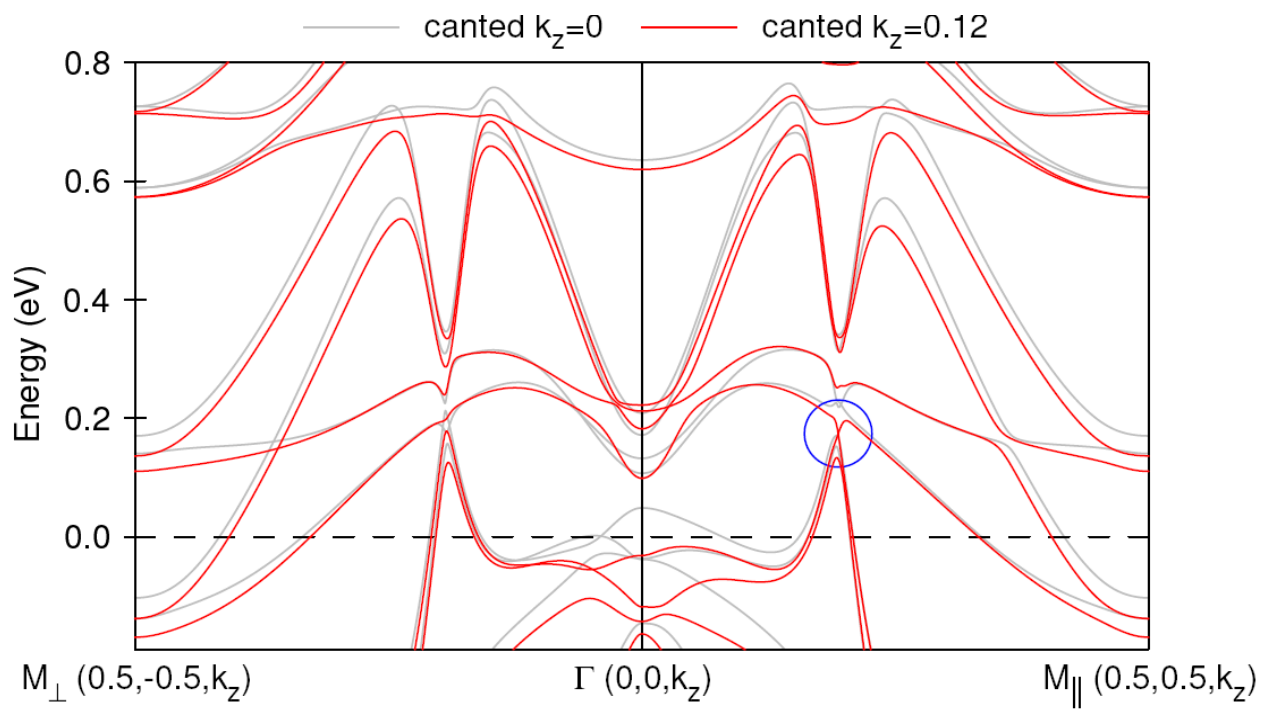
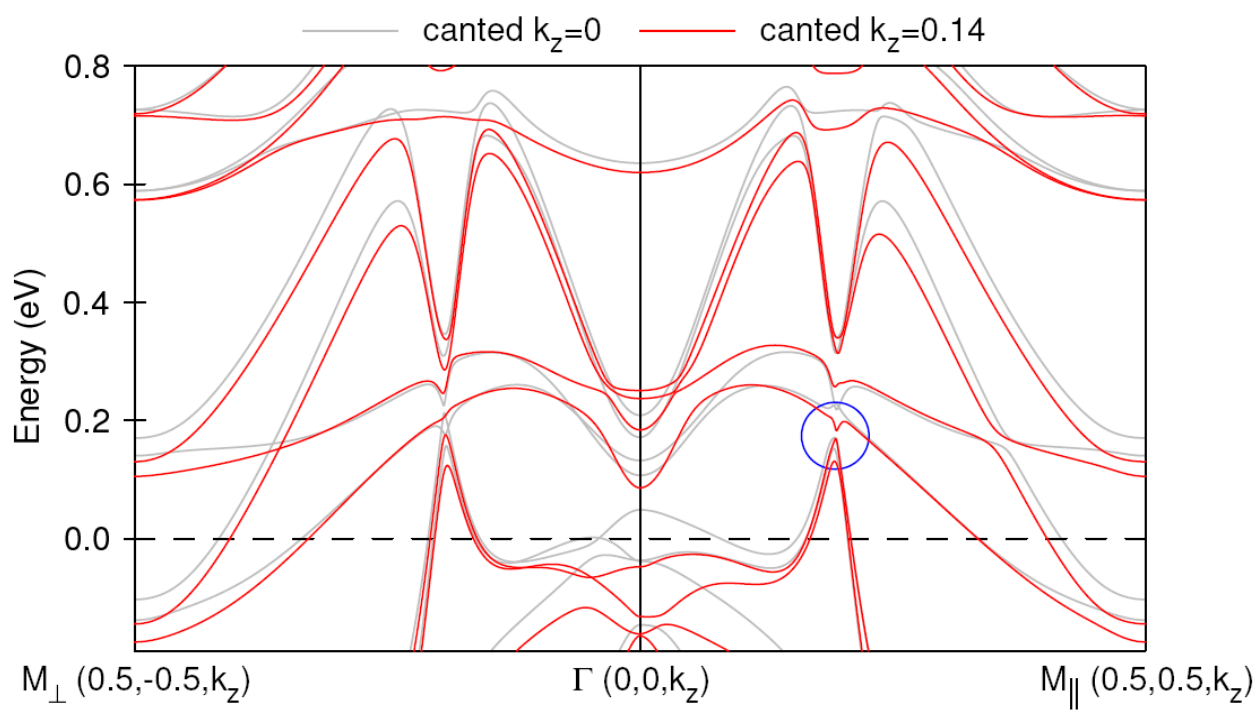

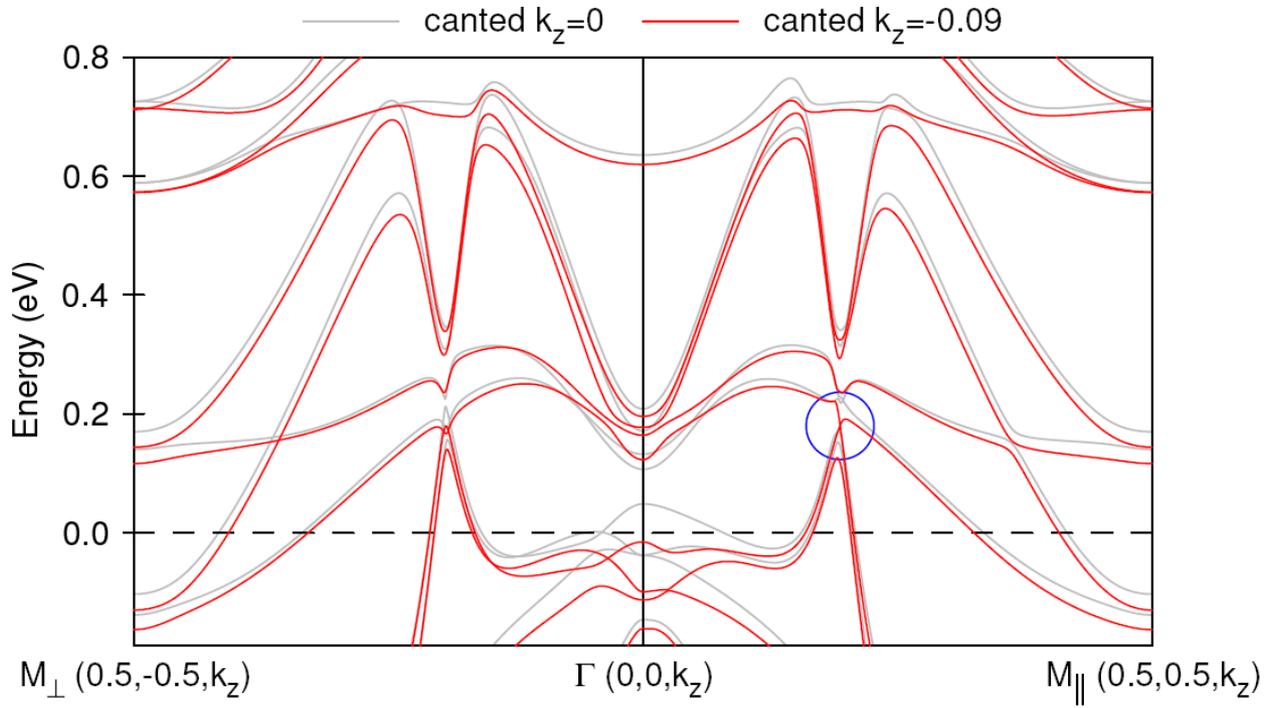

**Figure S6.** Weyl crossings in high symmetry planes of YbMnBi$_2$. $\Gamma M_{\parallel}$ and $\Gamma M_{\perp}$ are directions along and perpendicular to the net magnetization respectively.

As is seen from the calculations, there are two closely separated crossings of the singly degenerate bands on the $\Gamma M_{\perp}$ direction, but these are not 3D Weyl points, as evolution with kz shows. These crossings make a loop in the vertical plane. These two symmetry related loops are shown in Fig. 2c. On the $\Gamma M_{\parallel}$ direction, in contrast, a pair of true 3D Weyl points is observed at kz=0.12 and kz=-0.09. Away from vertical Z$\Gamma$M planes all crossings become avoided. The locations of 3D Weyl points are symmetrical with respect to earlier detected 3D Dirac point (kz=0.015) for collinear configuration of spins, implying that this pair of Weyl points is created by the time-reversal symmetry breaking (canting). In order to test this, we have carried out the calculations for different values of canting angle. Indeed, the distance between the Weyl point and initial 3D Dirac point decreased with decreasing the canting angle (Table S1).

**Table S1.** Coordinates of the Weyl point as a function of canting angle.

| Canting angle (°) | $k_x$ ($2\pi/a$) | $k_{z+}$ ($2\pi/c$) |
|---|---|---|
| 10 | 0.194 | 0.120 |
| 5 | 0.193 | 0.070 |
| 2 | 0.193 | 0.040 |
| 1 | 0.193 | 0.027 |
| 0 | 0.193 | 0.015 |

We have also scanned other portions of BZ and found another set of Weyl points, lying away from high-symmetry planes and directions. First, in Fig. S7 we show the radial cuts from $\Gamma$-point towards the Fermi surface contour for kz=0 and kz=$\pi$/c.

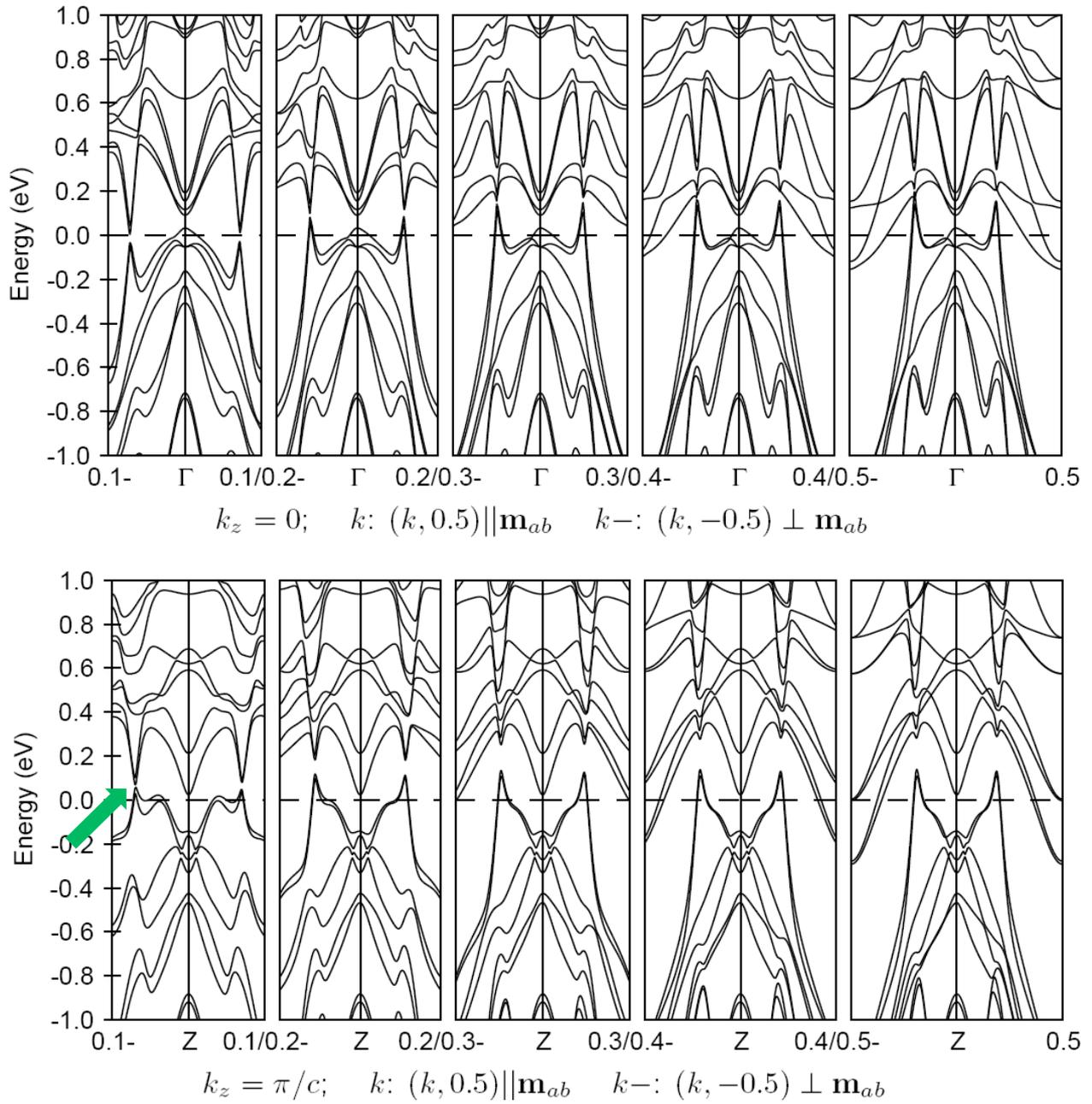

**Figure S7.** YbMnBi$_2$: canting with m$_{ab}$||(0.5, 0.5, 0). Green arrow shows the location where the gap is very small but finite (~8 meV).

As is seen, the very small gap is detected in the ZAR-plane signaling the proximity to a Weyl point. Further scanning of momentum space resulted in identification of Weyl points with very anisotropic dispersions at kz=0.131 (see Fig. S8). The Weyl point is found at (0.394,0.045, 0.131). The kx and ky of this position corresponds to the point where the hole-like lens is connected to electron-like pocket. These results are presented in Fig. S9.

$k_z = 0.12$

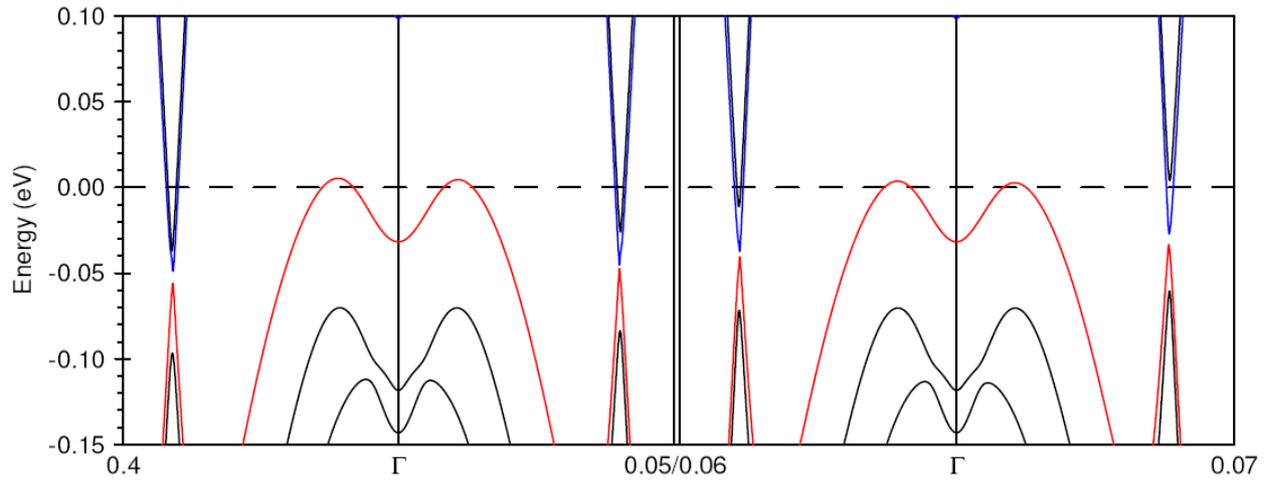

$k_z = 0.13$

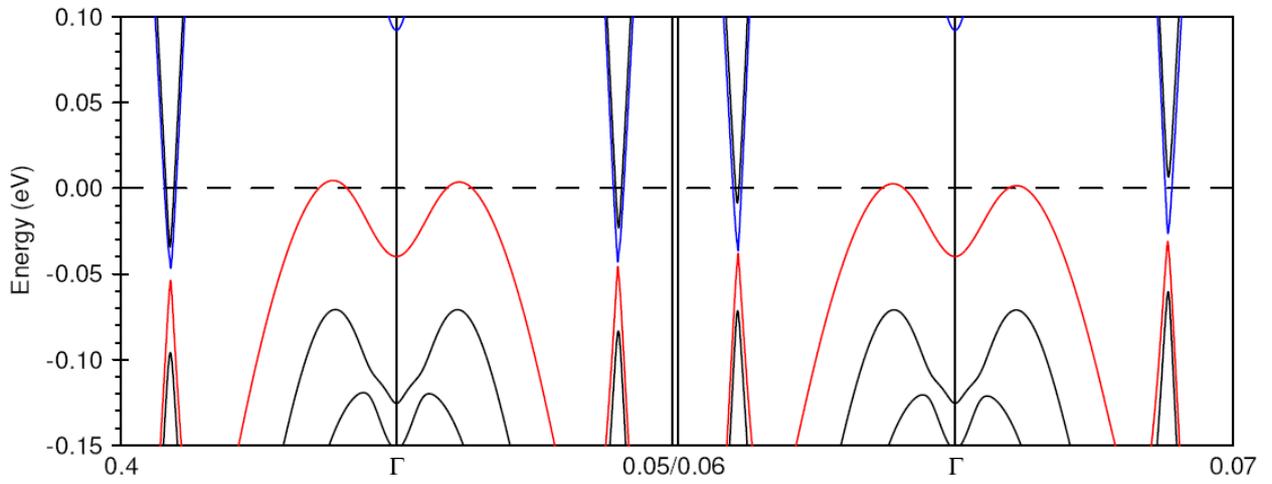

$k_z = 0.14$

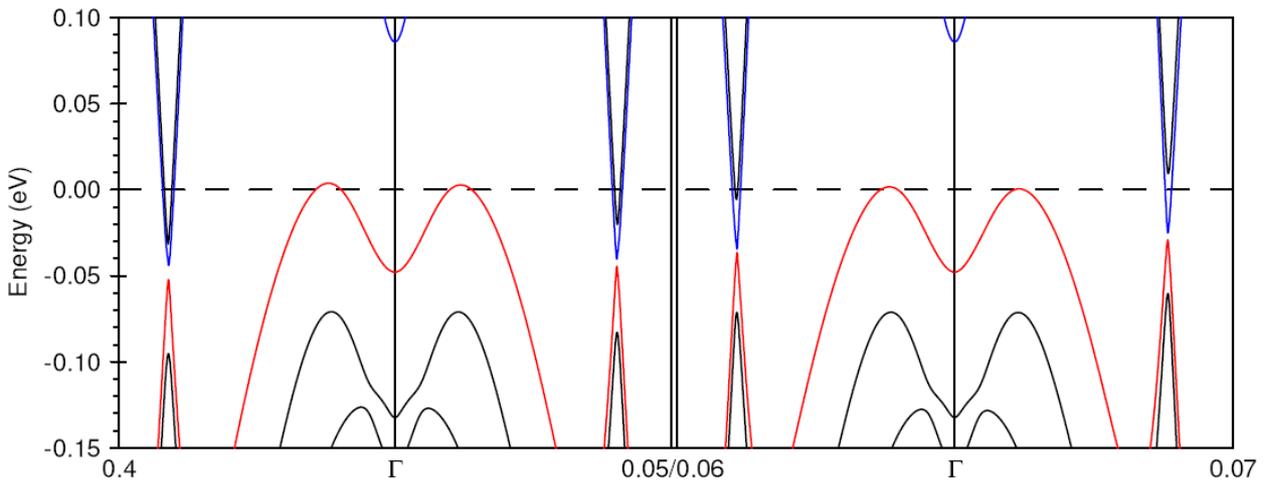

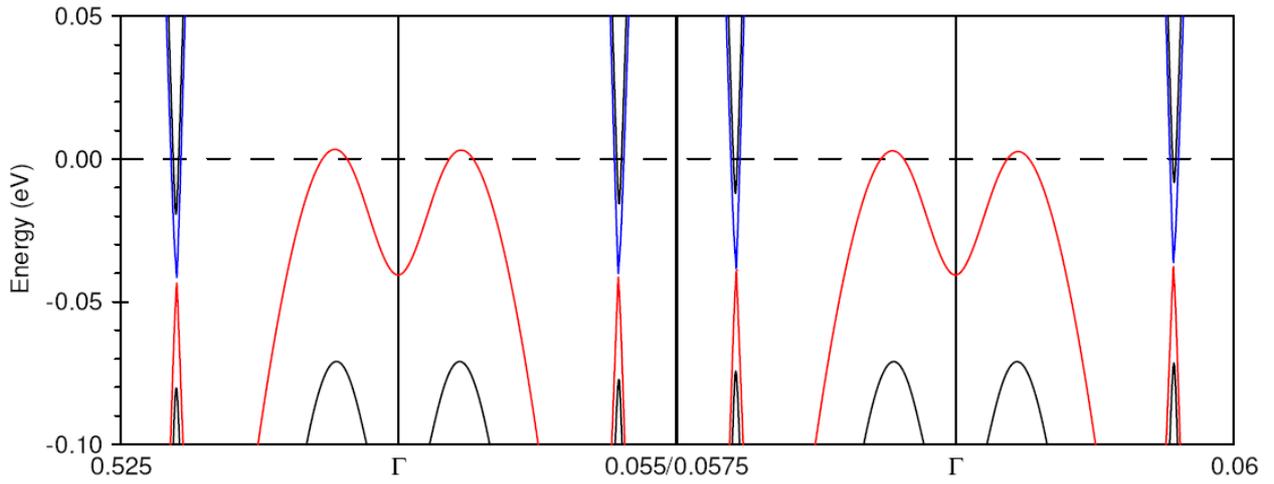

**Figure S8.** Identification of the Weyl points in YbMnBi2 corresponding to the experiment. Only the last panel shows the Weyl point itself when going from Γ to (0.0575, 0.5, 0.131). Other directions show very small gaps indicating a very anisotropic behavior close to the Weyl node.

Other symmetry related Weyl points are at (0.394,-0.045,0.131), (0.045, 0.394,0.131) and (-0.045, -0.394,0.131). As follows from the presented data, there are two $k_z$ values which define the $k_x$-$k_y$ planes where Fermi surface may look continuous in the experiment: $k_z=\pi/c$ where the gaps are very small and $k_z$~0.1 where the true 3D Weyl points are observed.

EXPERIMENT

In order to understand the complete 3D electronic structure of both materials we have recorded Fermi surface maps at different excitation photon energies. The results for EuMnBi$_2$ are shown in Fig. S9. Four lenses Fermi surface remains visible at different photon energies, which is not surprising since it originates from the 2D networks of Bi atoms. On the other hand, it can be due to the probing similar $k_z$s. As is seen from two middle panels, intensity distribution at the Fermi level only slightly changes as a function of light polarization. Intensity near the Γ-point does change and represents 3D band which also cross the Fermi level at particular $k_z$.

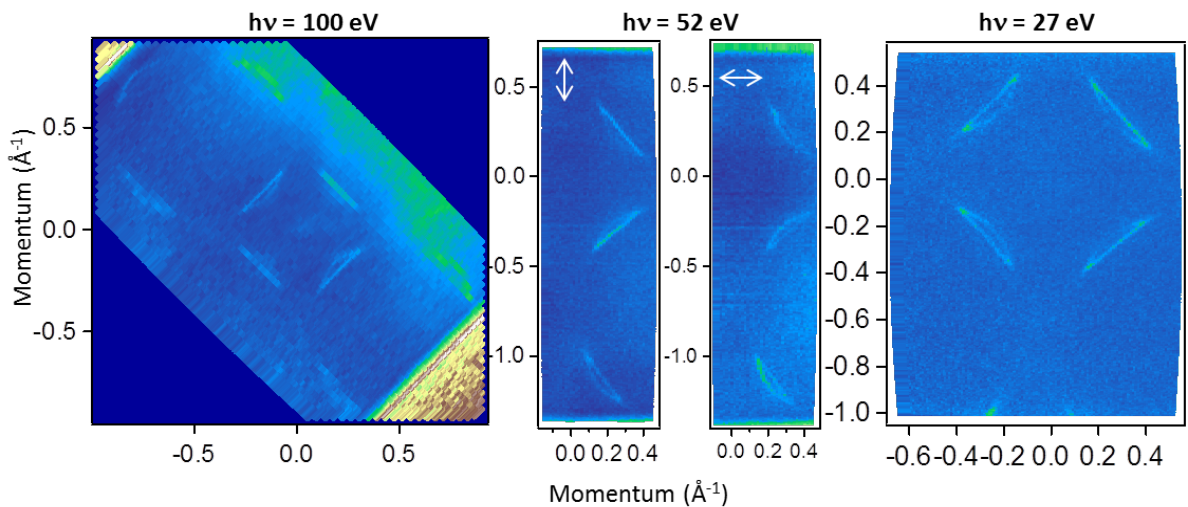

**Figure S9.** Fermi surface maps of EuMnBi$_2$ for different excitation energies. White arrows indicate the direction of light polarization.

Since EuMnBi$_2$ did not show any evidence for the 3D Dirac or Weyl points n the following we deal with YbMnBi$_2$ only. In spite of very pronounced two-dimensionality of the features originated from the Bi-networks, we were able to detect clearly periodic patterns of intensity distribution along ΓM direction (Fig. S10a,b) and determine photon energies which correspond to Γ and Z points in YbMnBi$_2$ (Table S2).

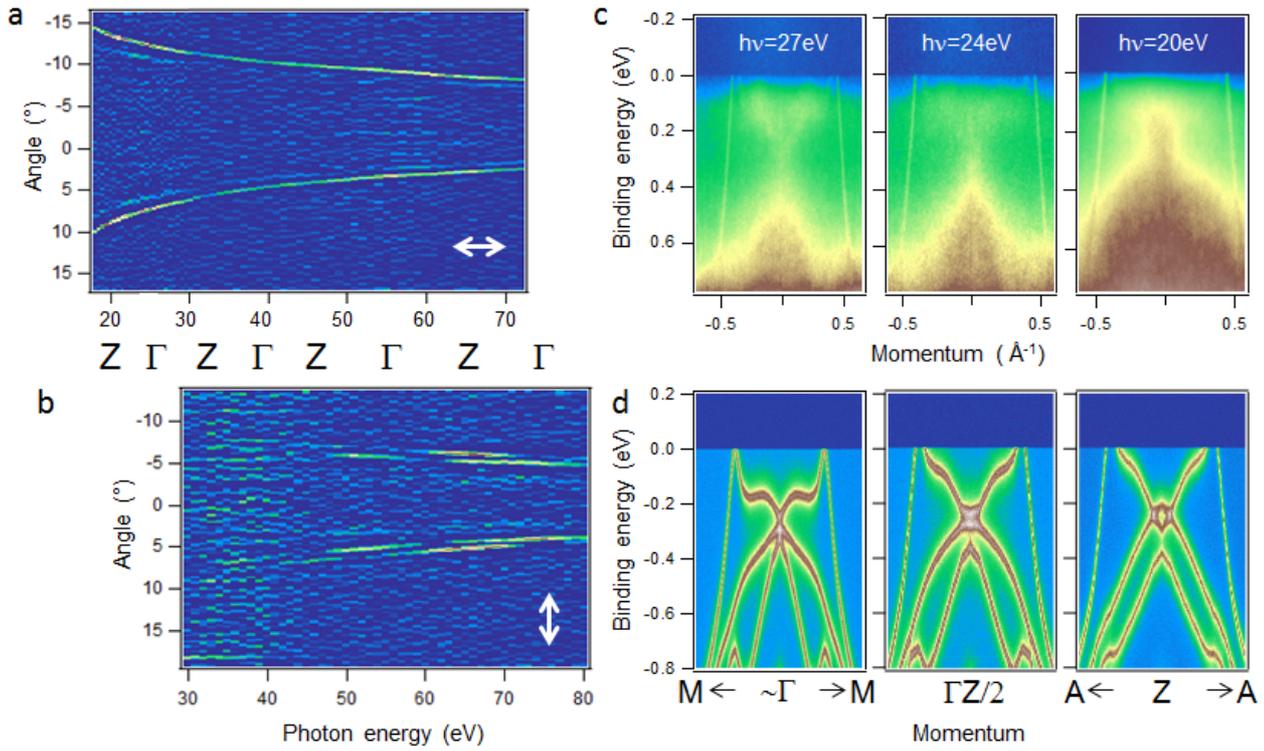

**Figure S10.** YbMnBi$_2$: a) and b) second derivatives of the photon energy dependence of the integrated (40 meV) spectral weight close to the Fermi level for s- and p-experimental geometries along ΓM-direction. c) Examples of intensity maps taken at 27 (~Γ), 24 (~ΓZ/2) and 20 eV (~Z). d) Corresponding calculated bands. Fermi level is adjusted for better agreement with the experiment. The evolution of the 3D band with kz is seen.

**Table S2.** Photon energies at which high symmetry points are accessible considering that inner potential is 6.5 eV.

| Symmetry point | Photon energy (eV) | | | |
|---|---|---|---|---|
| Γ | 26 | 39 | 56 | 75 |
| Z | 20 | 32 | 47 | 65 |

Intensity distributions at the energies corresponding to high-symmetry points are compared with calculations in panels c and d of Fig.S10. There is a qualitative agreement between two datasets confirming our assignment made in Table S2 and defined by the value of inner potential. We have recorded Fermi surface maps using different photon energies also for YbMnBi$_2$. The results are shown in Fig.S11.

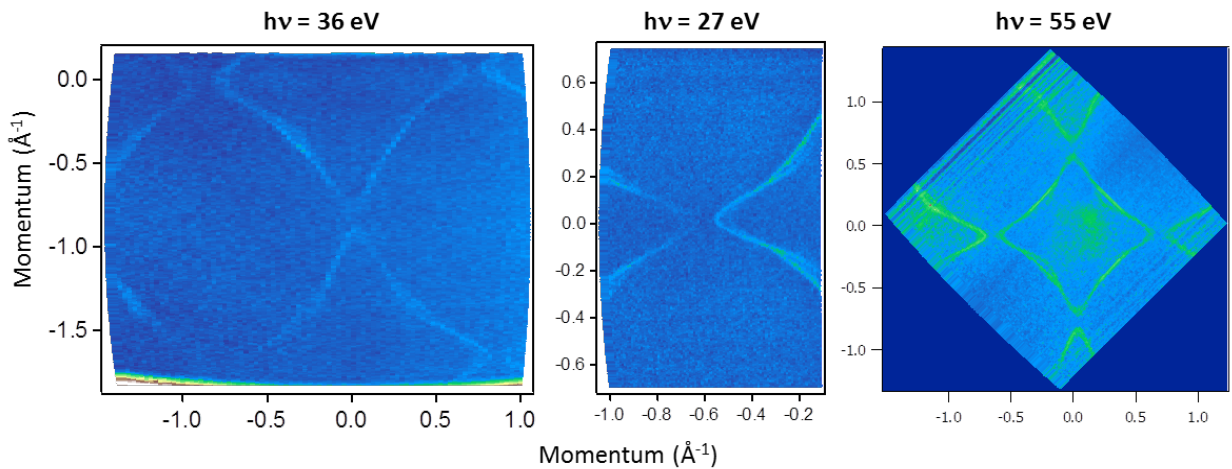

**Figure S11.** Fermi surface maps of YbMnBi$_2$ for different excitation energies.

We emphasize here that the energies have been selected before the value of the inner potential has been found, i.e. before the assignment to G and Z points. The only criterion for the selection was the well-defined photoemission signal and sharpness of the features. At other photon energies we were not able to obtain a clear picture of the Fermi surface. Now it can be explained in terms of two regions of kz identified above: at those kz which correspond to either minimal gaps or Weyl crossings the kz dispersion is naturally weaker. Fine structure of the lenses and arcs is not clearly distinguishable in these maps since their resolution requires better momentum resolution which is achievable either using lower photon energies or particular geometry. In Fig. S12 we show such a case. The sample has been oriented very precisely and the Fermi surface map has been recorded with the very small step in angle. One can see that the lenses became more separated from the arcs.

We have also checked that the lifting of the degeneracy is observed at different photon energies to exclude its artificial origin. Corresponding data sets are shown in Fig. S13. It is seen that the linear features dispersing from higher binding energies split and only one pair reaches the Fermi level for both, positive and negative angles.

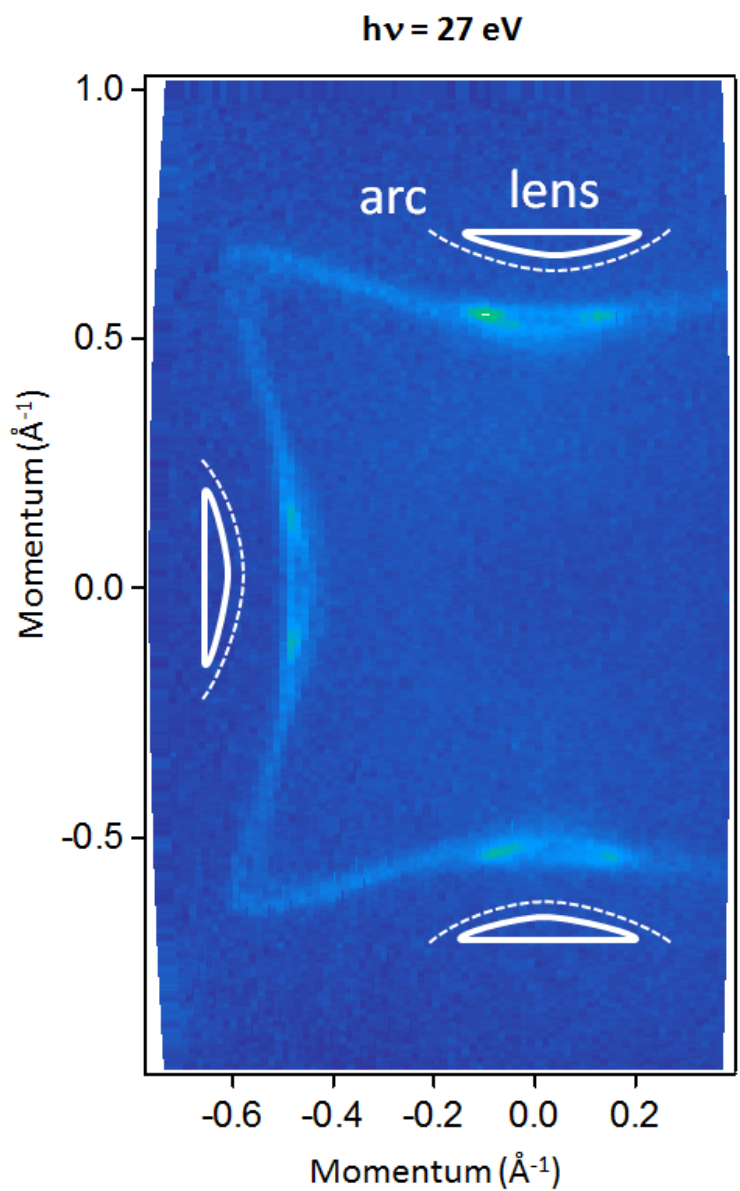

**Figure S12.** Fermi surface map of YbMnBi$_2$ taken in particular geometry to distinguish the lenses from arcs. White lines are schematic guides to eye shifted from the experimental features.

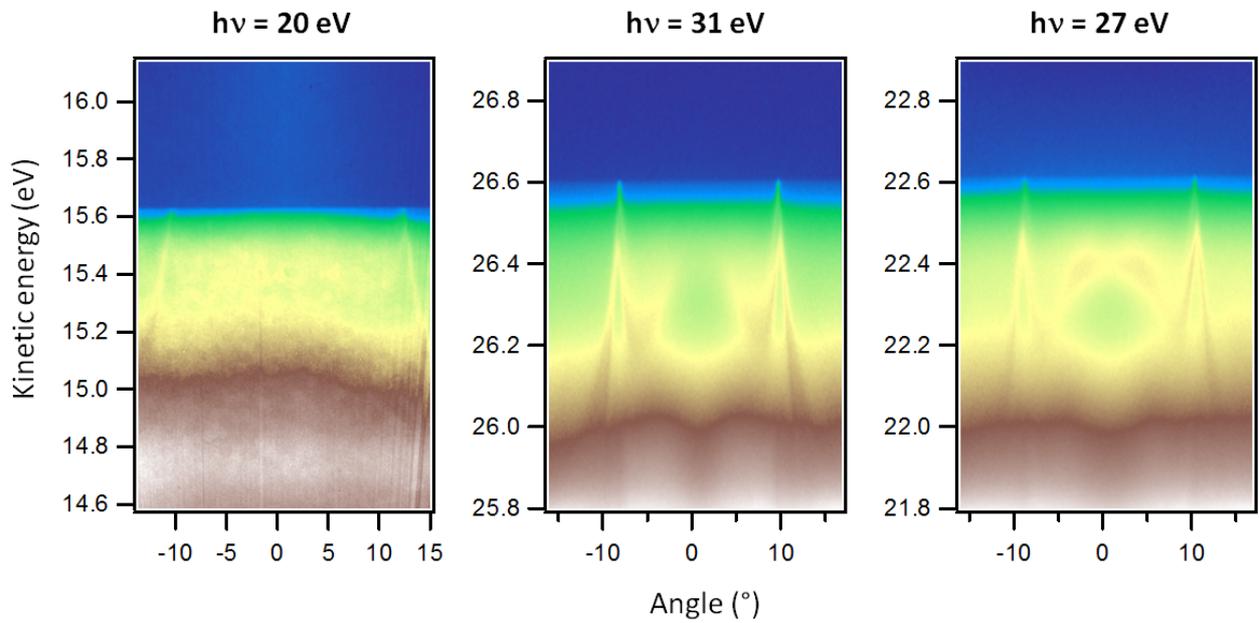

**Figure S13.** YbMnBi$_2$: Momentum cuts through or close to the Weyl points at different photon energies.

Again, this behavior can be understood considering the band structure calculations taking into account canted antiferromagnetism of Mn atoms: the photon energies at which we saw the splitting approximately correspond to the discussed above two values of kz, where kz dispersion is smaller.

In order to confirm that the lens is supported by the non-degenerate dispersion, we have studied the width of momentum distribution curves in the immediate vicinity of the Fermi level and approached the base of the lens at a right angle to ensure the highest possible angular resolution (Fig. S14).

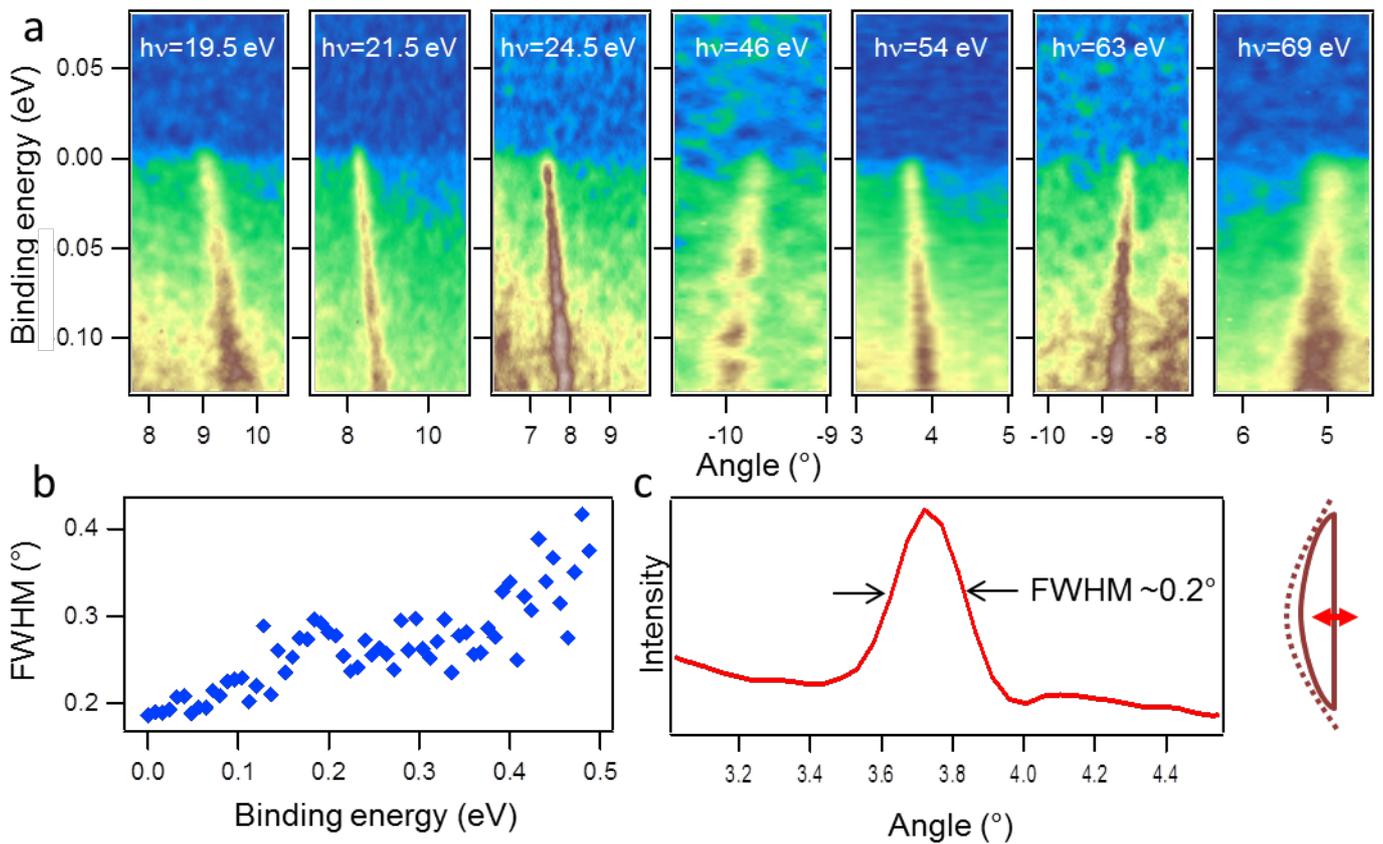

**Figure S14.** YbMnBi$_2$: zoom in to the dispersion supporting the base of the lens as shown in the inset. a) photon energy dependence, b) typical FWHM of the k$_F$-MDC. c) typical k$_F$-MDC.

In Fig. S14a we plot these dispersions measured at different photon energies. As follows from the data, there is always a single feature and its width in momentum is resolution limited corresponding to ~0.2° (Fig. S14c). Only at higher binding energies (~150 meV) the FWHM becomes larger, possibly indicating the presence of another feature. Indeed, as is seen from Fig. S15, approximately at these energies one can see the dispersion of another split band directly.

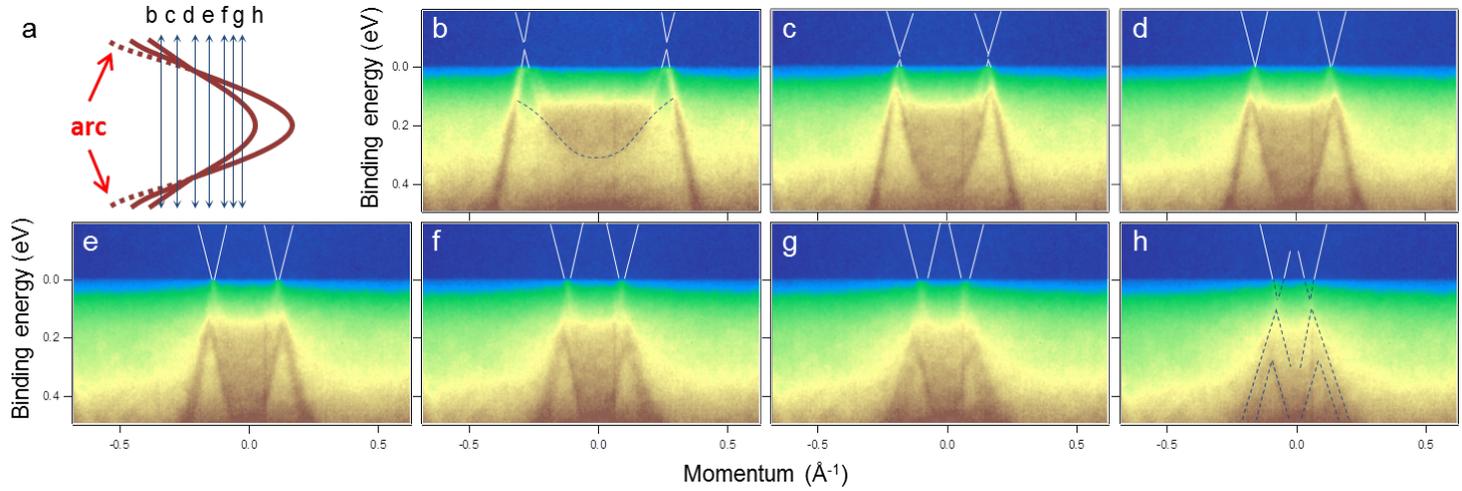

**Figure S15.** YbMnBi$_2$: data taken at highest possible in our experiment energy and momentum resolutions (hν=20 eV). Dashed and white lines are guides to the eye to help to track the dispersion of the bulk states. Surface Fermi arc is visible in panel b.

Our data indicate that both hole-like lens and electron-like pocket are supported by a single dispersion. The other magnetically split band remains at higher binding energies and does not cross the Fermi level. Figure S15 also demonstrates that YbMnBi$_2$ is Weyl semimetal of second type, i.e. the Weyl point is situated where the hole-like lens touches the electron-like pocket.

Finally, we present the structural data for the novel Weyl semimetal of type II - YbMnBi$_2$ – in Table S3.

**Table S3**: Single crystal structure determination for YbMnBi$_2$, experimental data taken at 100 K.

| Phase | YbMnBi$_2$ |
|---|---|
| Symmetry | Tetragonal, *P4/nmm* (No. 129) |
| Cell Parameters (Å) | a = 4.478(1), c = 10.819(2) |
| | α = β = ϒ = 90° |
| Wavelength (Å) | Mo Kα = 0.7107 |
| V (Å$^3$) | 216.94(3) |
| Z | 1 |
| Calculated Density (g cm$^{-3}$) | 4.94(1) |
| Formula Weight (g mol$^{-1}$) | 645.9 |
| Absorption Coefficient (mm$^{-1}$) | 52.42 |
| F$_{000}$ | 261.0 |
| Independent/Observed Reflections | 194/183 |
| Data/Restraints/Parameters | 194 / 0 / 23 |
| Difference e- density (e/ Å$^3$) | +7.85 to -3.45 |
| $R_1$ (all reflections) | 0.044 |
| $R_1$ Fo > 2σ(Fo)) | 0.0423 |
| $wR_2$ | 0.108 |
| $R_{int}/R(\sigma)$ | 0.0542/0.0304 |
| GooF | 1.34 |